\documentclass[twocolumn]{aastex631}

\usepackage{graphicx}	
\usepackage{amsmath}	
\usepackage{amssymb}	
\usepackage{comment}
\usepackage{booktabs}  
\usepackage[hyphens]{url} 
\usepackage[starfontserif]{starfont}
\DeclareSymbolFont{starfontsym}{OT1}{sts}{m}{n}
\DeclareMathSymbol{\MathNeptune}{\mathord}{starfontsym}{71}
\usepackage{upgreek}

\shorttitle{Water on the hot Neptune HD~219666~b}
\shortauthors{Murphy, Beatty, Welbanks, \& Fu}

\begin{document}

\title{HST Transmission Spectra of the Hot-Neptune HD~219666~b: Detection of Water and the Challenge of Constraining Both Water and Methane}

\correspondingauthor{Matthew M. Murphy}
\email{mmmurphy@arizona.edu}

\author[0000-0002-8517-8857]{Matthew M. Murphy}
\affiliation{Steward Observatory, 933 N. Cherry Avenue, Tucson, AZ 85719, USA}

\author[0000-0002-9539-4203]{Thomas G. Beatty}
\affiliation{Department of Astronomy, University of Wisconsin--Madison, Madison, WI 53703, USA}

\author[0000-0003-0156-4564]{Luis Welbanks}
\thanks{51 Pegasi b Fellow}
\affiliation{School of Earth and Space Exploration, Arizona State University, Tempe, AZ, USA}

\author[0000-0002-3263-2251]{Guangwei Fu}
\affiliation{Department of Physics and Astronomy, Johns Hopkins University, Baltimore, MD, USA}

\begin{abstract}
Although Neptunian-sized (2 - 5 R$_{\earth}$) planets appear to be extremely common in the Galaxy, many mysteries remain about their overall nature. To date, only 11 Neptunian-sized planets have had their atmospheres spectroscopically characterized, and these observations hint at interesting diversity within this class of planets. Much of our understanding of these worlds and others derive from transmission spectroscopy with the Hubble Space Telescope's Wide Field Camera 3 (HST/WFC3). One key outcome of HST/WFC3 observations has been the consistent detection of water but no methane in Neptunian atmospheres, though recent James Webb Space Telescope (JWST) observations are potentially starting to overturn this ``missing methane" paradigm. In this work, we present the transmission spectrum of the hot Neptune HD~219666~b from 1.1 - 1.6 $\mu$m from two transit observations using HST/WFC3 G141. Our fiducial atmospheric retrieval detects water at $\sim$3-$\sigma$ in HD~219666~b's atmosphere and prefers no contribution from methane, similar to these previous observations of other planets. Motivated by recent detections of methane in Neptunian atmospheres by JWST, we explore additional models and find that a methane-only scenario could adequately fit the data, though it is not preferred and likely unphysical. We discuss the impact of this methane detection challenge on our understanding of planetary atmospheres based on HST/WFC3 observations alone, and where JWST observations offer a solution. 
\end{abstract}


\section{Introduction}
\label{sec:intro}

As of this writing, 5,839 planets have been confirmed outside of the Solar System (NASA Exoplanet Archive). Driven by dedicated surveys from both the ground and space, this number is steadily increasing. One of the unexpected results from these discovery campaigns is that there exists a large population of planets between approximately 2 - 5~R$_{\earth}$ or 0.5 - 1.3~R$_{\MathNeptune}$ in size. These Neptunian-sized planets make up $\sim$42\% of the current transiting exoplanet population (NASA Exoplanet Archive), and may even be the most common type of planet in the Galaxy \citep{fulton18_demographics, hsu19_demographics}. We intentionally use the term ``Neptunian" here to broadly refer to both sub-Neptune-sized and Neptune-sized exoplanets, which may be compositionally similar, as ``Neptune-sized" may exclude the former class. Understanding these Neptunian-sized worlds is therefore essential to understanding the nature of the planetary population as a whole. 

Despite their prevalence, only around a dozen Neptunian-sized worlds have been spectroscopically characterized: K2-18 b \citep{benneke19_k218b, tsiaras2019_k218b, madhusudhan2023_k218b, wogan2024_k218b}, GJ 3470 b \citep{ehrenreich2014_gj3470, benneke2019_gj3470}, GJ 1214 b \citep{berta2012_gj1214b, kreidberg2014_gj1214b, kempton2023_gj1214b, schlawin2024_gj1214b, ohno2024_gj1214b}, TOI-270~d \citep{mikalevans23_toi270d, benneke24_toi270d, holmberg24_toi270d}, GJ 436 b \citep{knutson2014_gj436b}, HD 97658 b \citep{knutson2014_hd97658, guo2020_hd97658b}, HD~3167~c \citep{mikalevans2021_hd3167c}, HD~106315~c \citep{kreidberg2022_hd106315c}, GJ~9827~d \citep{roy2023_gj9827d}, LTT~9779~b \citep{crossfield2020_ltt9779b, dragomir2020_ltt9779b, edwards2023_ltt9779b, radica2024_ltt9779b}, and HAT-P-11 b \citep{fraine2014_hatp11b, chachan2019_hatp11b}. These observations already point to immense diversity within this class of planets though, with inferred atmospheric metallicities spanning several orders of magnitude times the solar value, and transmission features ranging from strong \citep[e.g.,][]{fraine2014_hatp11b} to completely muted \citep[e.g.,][]{kreidberg2014_gj1214b}. Continued spectroscopic characterization is the only way to unravel this diversity, and understand how these planets form and evolve. 

Just as important as taking spectroscopic observations is properly interpreting them, and understanding the capabilities and limitations of each part of our analyses -- whether that is the data or the models. A key recent example of this was the so-called ``missing methane" problem, where observations of cool ($\lesssim$1000~K) exoplanetary atmospheres, including the aforementioned planets, with the Hubble Space Telescope's Wide Field Camera 3 (HST/WFC3) consistently did not detect methane. Methane should be the primary carbon reservoir in these cool atmospheres and the HST/WFC3 G141 bandpass probes several strong methane absorption bands, so these nondetections challenged the community for some time. Since the advent of follow-up observations with the James Webb Space Telescope (JWST), however, this picture has changed. Using JWST, methane has now been detected in the atmospheres of several Neptunian-sized (2 - 5 R$_{\earth}$) worlds including K2-18~b \citep{madhusudhan2023_k218b, wogan2024_k218b}, GJ~3470~b \citep{beatty24_gj3470b}, and TOI-270~d \citep{benneke24_toi270d, holmberg24_toi270d}; updating previous nondetections from HST/WFC3. 

One important outcome of these HST/WFC3 nondetections was the development of theories of chemical disequilibrium, which acts to deplete the atmospheric abundance of methane relative to that expected under chemical equilibrium. Disequilibrium processes, including vertical mixing and tidal heating, became the leading hypotheses to explain the HST/WFC3 nondetections. In fact, the abundances of methane inferred by JWST thus far are still consistent with those predicted by chemical disequilibrium \citep[e.g.,][]{welbanks24_wasp107}, suggesting it is indeed a widespread process in exoplanetary atmospheres. Nevertheless, these recent JWST detections of methane raise the question of why HST/WFC3 observations consistently missed the methane, and whether our population-level knowledge of exoplanetary atmospheres based on HST/WFC3 results needs to be revisited. 

To begin exploring these questions, in this work we present spectroscopic observations of the hot Neptune HD~219666~b using HST/WFC3 G141. One of the first exoplanets discovered by the Transiting Exoplanet Survey Satellite (TESS) mission \citep{esposito2019}, HD~219666~b is similar to Neptune in size (R$_{p}$ = 4.79~R$_\earth$ = 1.24~R$_\MathNeptune$) and mass (M$_{p}$ = 17.6~M$_\earth$ = 1.02~M$_\MathNeptune$). HD~219666~b is on a 6.036~day orbit around an old ($\sim$10~Gyr) Sun-like star, subjecting it to an equilibrium temperature of approximately 1080~K and making it the second hottest Neptunian planet to be characterized to date. The hottest to be spectroscopically characterized thus far is LTT-9779~b \citep{crossfield2020_ltt9779b, dragomir2020_ltt9779b, edwards2023_ltt9779b, radica2024_ltt9779b} but, unlike LTT-9779~b (T$_{eq}$ $\sim$ 2000~K), HD~219666~b exists at a temperature in the regime where methane should contribute appreciably to the atmosphere \citep{moses2013_eqabundances}. 

\section{Observations \& Data Reduction}
\label{sec:obs}

\subsection{Observation Information}
\label{subsec:obs_info}

We observed two transits of HD~219666~b with the HST, one each from programs 15698 (PI: T. Beatty) and 15969 (PI: G. Fu). Both observations used the IR channel of the WFC3 with the G141 grism in spatial scan mode, providing slitless spectroscopy between approximately 1.1 to 1.7~$\mu$m. We supplemented these HST data with 12 additional transit observations of HD~219666~b from the TESS, which are publicly available from the Mikulski Archive for Space Telescopes\footnote{\url{https://archive.stsci.edu/}}, as well as radial velocity (RV) observations from the HARPS spectrograph as presented and tabulated by \cite{esposito2019}, and stellar magnitude measurements from \cite{hog2000Bmagnitude}, \cite{2mass2003}, and \cite{wise2014}. 

In HST program 15698 (PI: T. Beatty), we originally planned to observe three total transits of HD~219666~b across three visits. However, the first two visits, on UT 2019 June 2 and 2019 August 7, missed the planet's transit due to an incorrect ephemeris. Therefore, in this work, we only present data from the program's third visit, on UT 2019 November 12. As mentioned, this observation used the WFC3 IR camera in spatial scan mode, doing round-trip scanning at a rate of 0.215 arcsec/s. At the beginning of each consecutive orbit of HST around the Earth, we took a direct image of HD~219666 using the F126N filter for wavelength calibration. Afterward, we switched to the G141 grism and took seventeen spectroscopic exposures during that orbit's visibility window with a scan duration of $\sim$111 s each in MULTIACCUM mode using SPARS10 with NSAMP = 15. The reduction of this data is described further in Section~\ref{subsec:imagereduction}. We note this observation was also independently analyzed in the population study of \cite{edwards2023_popstudy}.

In HST program 15969 (PI: G. Fu), we also originally planned three visits of the HD~219666 system using G141. The first visit on UT 2019 November 18 also missed the transit due to an incorrect ephemeris, and the final visit on UT 2020 June 28 only captured a partial transit and exhibited very strong systematics. In this work, we present only the data from this program's second visit on UT 2020 May 5. Unlike program 15698, in this observation we only took a single direct image of HD~219666 at the start of the visit, also using the F126N filter. Then, we took 44 exposures per HST orbit with scan durations of 81 seconds (scan rate of 0.216 arcsec/s) in MULTIACCUM mode with NSAMP = 12, again doing round-trip scanning. Due to a more favorable sky position of HD~219666 at this time, this visit more completely sampled HD~219666~b's transit than that of program 15698. We reduced this data using the same method as that for program 15698, described further in Section~\ref{subsec:imagereduction}. Hereafter, we will refer to the transit from program 15698 as ``HST Visit 1" and the transit from 15969 as ``HST Visit 2".

TESS captured four transits of HD~219666~b in 2018, four more in 2020, and another four in 2023. Each transit was observed at a 120~second cadence. We downloaded and used the \texttt{light curve} files from the default TESS reduction pipeline, which successfully removed all obvious, significant systematic trends in the light curves. To expedite our analysis, we cut out all excess baseline measurements between consecutive transits, and focused only on data within 5~hr of each individual transit's midpoint. 

\cite{esposito2019} measured the radial velocities of HD 219666 over 1 month in 2018 with the HARPS spectrograph, and tabulated their measurements in their Table~3. We took these tabulated measurements as given for inclusion in our analysis. 

In addition, HD~219666 has been observed during a variety of imaging surveys, providing a set of stellar magnitude measurements which we collected from the NASA Exoplanet Archive. Described further in Section~\ref{subsec:methods_LCfitting}, we used these TESS transit, RV, and stellar magnitude data to precisely constrain the physical and orbital properties of HD~219666~b to support our analysis and interpretation of the HST observations. 

\subsection{HST/WFC3 Image Reduction}
\label{subsec:imagereduction}

We first used each direct image of HD 219666 to compute a wavelength solution for all spectral exposures in that orbit (for the first visit) or whole visit (for the second visit). HD~219666 was the only star visible in each image. We determined its image coordinates using DAOStarFinder \citep{DAOfinder}, then converted these to the corresponding physical position on the WFC3 detector using the specified subarray x- and y-offsets (\texttt{LTV1} and \texttt{LTV2} in the image headers). Then, we calculated the pixel-to-wavelength solution using the calibrated measurements of \cite{2016wfcwavelengthcalibrationreport}. 

After computing the wavelength solution, we began reducing each spectral exposure. We separated the sequential up-the-ramp read-outs in each \texttt{MULTIACCUM} exposure, splitting each exposure into multiple subexposure images. Our custom reduction pipeline is structured to hierarchically treat each subexposure in an exposure, then each exposure in an HST orbit, for all orbits in a single visit. We calculated the time stamp of each subexposure as half the sampling time (header value \texttt{SAMPTIME}) subtracted from the UT time stamp at the end of read-out (header value \texttt{ROUTTIME}), then used \texttt{astropy}'s \texttt{time} and \texttt{coordinates} modules to first convert this to \texttt{JD TDB} format, then adjusted for the relative light travel time to Earth to convert into \texttt{BJD TDB} format. Next, we performed background subtraction on each subexposure. We masked out the science spectrum, computed the median value of the unmasked background pixels, and subtracted this value from each pixel in the entire image. The mean background flux level was 11.5 electrons per second, with a standard deviation of 0.96 electrons per second during HST Visit 1, and 19.31/2.34 electrons per second during HST Visit 2. Following background subtraction, we flat-fielded each subexposure using the most recent most recent WFC3/IR G141 flat-field template provided by STScI\footnote{\url{https://www.stsci.edu/hst/instrumentation/wfc3/documentation/grism-resources/wfc3-g141-calibrations}} and the wavelength solution computed previously. HST/WFC3 does not scan perfectly vertically down the detector, so the spectrum shifts slightly along the dispersion direction over the course of an exposure which must be accounted for when flat-fielding as well as in later reduction steps. We estimated this shift as a function of time, treating it as a linear function of the side angle (header value \texttt{ANG\_SIDE}) and scan rate (header value \texttt{SCAN\_RAT}). After flat-fielding, we performed bad-pixel correction on each subexposure. We created a smoothed spectrum by convolving the image, row by row, with a 1D Gaussian function and then subtracted this from the original image. We defined bad pixels as those whose difference is greater than 4 times the corresponding row's standard deviation, and replaced their value with the median value of its neighboring pixels. 

After background subtraction, flat-fielding, and bad-pixel correction, we computed the trace of each subexposure's spectrum. Column-by-column, we computed the center of the spectrum (along the y-direction) by taking a flux-weighted average of the column. Then, we fitted these centers as a function of x-pixel (i.e., dispersion-axis pixels) using a second-order polynomial, which smooths over any outliers. We then collapsed each subexposure into a ``1D" spectrum -- brightness as a function of dispersion-axis pixels -- by summing, for each column, all pixels within 15 pixels from the trace position. We deshifted each 1D spectrum along the dispersion direction by the amounts determined earlier, then summed each spectrum together (column-wise) into a single 1D spectrum for that exposure. 

From the time series of 1D spectra constructed above, we first created a broadband light curve for each visit by integrating over each 1D spectrum between 1.075~$\mu$m and 1.80~$\mu$m. Then, we created a set of fifteen spectral light curves by binning each spectrum into evenly sized bins of $\sim$0.0338~$\mu$m full width each, corresponding to a resolution of R$\sim$40. We normalized each light curve separately by scan direction, dividing by the visit-wide median for points of that scan direction. 

\section{Data Modeling}
\label{sec:models}

\subsection{Light Curve Models}
\label{subsec:lcmodels}

We modeled each transit light curve as the product of a transit model $T(t)$ for the astrophysical signal, and a systematics model $S(t)$ for nonastrophysical signals introduced by the instrument. For the transit model, we used the \texttt{BATMAN} python package \citep{kreidberg2015batman} with a quadratic limb-darkening prescription. For the systematics model, we fit a linear visit-long trend in brightness with time $r(t)$ to each individual light curve. We centered this linear trend around the visit's median time, so that
\begin{eqnarray}
r(t) = s_t \left( t - t_{\text{median}} \right) + b, \label{eqn:linearramp}
\end{eqnarray}
where $s_t$ is the slope and $b$ is the intercept. For the HST visits, we also fit an empirical model for the WFC3 detector charge-trapping effect \citep[see, e.g.,][]{long2013wfc3chargetrapping}, applied to each individual HST orbit as
\begin{eqnarray}
H(t) = A ~e^{- \frac{\left(t - t_o \right)}{\tau} }, \label{eqn:HSThook}
\end{eqnarray}
where $A$ is an amplitude, $t_o$ is the time when the corresponding HST orbit began, and $\tau$ is a timescale. Each orbit within a visit shared the same values for $A$ and $\tau$. All together, we modeled the relative flux light curves $F(t)$ for HST as $F_{\text{HST}}(t) = H(t)~ r(t)~ T(t)$ and for TESS as $F_{\text{TESS}}(t) = r(t)~ T(t)$. 

\subsection{Radial Velocity Model}
\label{subsec:rvmodel}

Following the method of \cite{exoplanetstext_RV}, we modeled the HD 219666 system's total RV signal as a function of time as
\begin{eqnarray}
v(t) = K \left[ \cos \left( \omega + f(t) \right) + e \cos \left( \omega \right) \right] + \gamma.
\end{eqnarray}
Here, $K$ is the RV semi-amplitude (i.e., the planet's contribution to the signal), $\omega$ is the planet's argument of periastron, $f(t)$ is the planet's true anomaly as a function of time, $e$ is the planet's eccentricity, and $\gamma$ is the bulk system radial velocity. Rather than treating the true anomaly as an independent variable, we calculated it at each time step using the planet's time of conjunction, orbital period, eccentricity, and argument of periastron, which were all fitting parameters (Section~\ref{subsec:methods_LCfitting}). 

\subsection{Stellar SED Model}
\label{subsec:sedmodel}

Table~\ref{tab:magnitudes} lists the measured magnitudes of HD~219666 in several bandpasses from \cite{hog2000Bmagnitude}, \cite{2mass2003}, and \cite{wise2014}. We used these magnitudes to fit HD~219666's spectral energy distribution (SED), in conjunction with the transit and RV measurements, to constrain the properties of both the star and planet. 

\begin{table} \centering
\caption{Literature measurements of HD~219666's apparent magnitude.}
\begin{tabular}{lcccl}\toprule
 Bandpass & Magnitude & Reference \\ 
\midrule 
B & 10.60 $\pm$ 0.03 & \cite{hog2000Bmagnitude}  \\
2MASS J & 8.56 $\pm$ 0.02 & \cite{2mass2003} \\
2MASS H & 8.25 $\pm$ 0.04 & \cite{2mass2003} \\
2MASS K & 8.16 $\pm$ 0.03 & \cite{2mass2003} \\
WISE1 & 8.08 $\pm$ 0.02 & \cite{wise2014} \\
WISE2 & 8.14 $\pm$ 0.02 & \cite{wise2014}
\\ \bottomrule 
\end{tabular}
\label{tab:magnitudes}
\end{table}


We first converted the magnitudes listed in Table~\ref{tab:magnitudes} to the corresponding band-integrated fluxes following \cite{cutri06_2MASScookbook} and \cite{cutri12_wisecookbook}. To infer the properties of HD~219666 via comparison to these fluxes, we downloaded a grid of precomputed stellar spectra for effective temperatures ($T_{eff}$) between 5000 and 6000 K, in steps of 100~K, with $log~g$ = 4.5 cm/s$^2$ and metallicity ([Fe/H] relative to solar) = 0 from the BT-Settl (CIFIST) stellar models \citep{BTsettl_1, BTsettl_2, BTsettl_3, BTsettl_4, BTsettl_5, BTsettl_6, BTsettl_7}. These parameters were the closest grid nodes to those inferred for HD~219666 by \cite{esposito2019}. We integrated each precomputed spectrum within the observed bandpasses to get the corresponding intrinsic model stellar flux, then linearly interpolated each bandpass' model flux along the temperature axis. To account for intrinsic dimming due to the distance between Earth and HD~219666, we multiplied each model flux value by $\left( R_\star / d_\star \right)^2$, where $R_\star$ is the radius of HD 219666 and $d_\star$ is the distance to it. In order to take advantage of precise prior measurements by the Gaia mission \citep{gaiadr22018}, we parameterized the distance to HD~219666 in terms of the parallax $\mu$, such that $d_\star = 1000 / \mu $ in units of parsecs, where $\mu$ is in units of milliarcseconds. To account for additional extinction due to interstellar dust, we also multiplied each model flux by $\exp \left( - B ~ A_V \right)$, where $B$ is the extinction base and $A_V$ is the amplitude of V-band extinction in magnitudes.

\subsection{Time Series and SED Model Fitting}
\label{subsec:methods_LCfitting}

Using the models described above, we fit the HST, TESS, and HARPS time series simultaneously via Bayesian inference using Markov Chain Monte Carlo (MCMC) sampling with the \texttt{emcee} python package \cite{emcee}. We followed the parameterization method of \cite{eastman2013parametrization}, parameterizing the orbital period $P$ as $\log_{10} \left( P \right)$, the semimajor axis $a$ as $\log_{10} \left( a / R_\star \right)$, the inclination $i$ as $\cos \left( i \right)$, and combined the eccentricity $e$ and argument of periapsis $\omega$ into the two new parameters $\sqrt{e} \cos{\left( \omega \right)}$ and $\sqrt{e} \sin{\left( \omega \right)}$. In total, we sampled these five parameters as well as the time of conjunction $t_c$, RV semi-amplitude $K$, system velocity $\gamma$, planet-to-star radius ratios ($R_p / R_\star$) in each bandpass, and the systematic model coefficients described in Section~\ref{subsec:lcmodels}. Shown later in Section~\ref{subsec:jointBBfit}, we found a significant offset in transit depth between the two HST visits, so we gave each visit its own $R_p/R_\star$ parameter. We left the quadratic limb-darkening coefficients free, with Gaussian-shaped Bayesian priors set to $u_1$ = 0.18 $\pm$ 0.01 and $u_2$ = 0.24 $\pm$ 0.01 for HST/WFC3 G141, and $u_1$ = 0.39 $\pm$ 0.05 and $u_2$ = 0.25 $\pm$ 0.05 for TESS. These limb-darkening coefficient priors are based on an ATLAS stellar model of HD~219666, calculated using \texttt{ExoCTK} \citep{ExoCTK}. Since existing literature measurements of HD 219666~b's orbital properties are based on the TESS data that we are refitting, we do not enforce any other priors. We sampled for 10,000 walker steps, in addition to a 1500 step initial burn-in, which was sufficient for all parameters to converge ($>$30x the mean autocorrelation time, $>$15x the largest autocorrelation time). 

We fit the stellar SED by itself since it does not explicitly share any parameters with the light curve and RV models. We again used the MCMC method, sampling the effective temperature $T_{eff}$, parallax $\mu$, stellar radius $R_\star$, and extinction $A_V$. We enforced Gaussian priors on the effective temperature and parallax of $T_{eff}$ = 5527 $\pm$ 65~K and $\mu$ = 10.62 $\pm$ 0.01 mas. We again ran this sampling for 10000 steps after a 1500 step initial burn-in, which was more than sufficient for convergence ($>$150x the largest autocorrelation time). 

When fitting the HST spectroscopic light curves, we fixed the orbital parameters to the best-fit values from our broadband fits described above, and fixed the quadratic limb-darkening coefficients in each wavelength channel to the corresponding ATLAS stellar model value. Again using MCMC, we fit each program's set of spectroscopic light curves separately, sampling just the $R_p/R_\star$ and systematic model coefficients in each spectral channel. We ran each channel's sampling for 6000 steps, burning the first 1000 of each ($>$70x the largest autocorrelation time). When fitting the HST light curves, in both the broadband and spectroscopic cases we neglected the first HST orbit's data as this orbit exhibits excess systematic effects and would not contribute significant additional scientific value. 

\subsection{Atmospheric Retrievals} \label{sec:methods_retrieval}

We explored HD~219666~b's atmospheric properties using flexible and agnostic forward models in a Bayesian inference procedure, known as free retrievals, to retrieve the chemical abundances of different gases, the vertical temperature structure, and the cloud/haze properties on the planetary atmosphere. This method does not assume any physicochemical equilibrium conditions, but instead aims to capture the atmospheric conditions directly through a series of parameters for the chemistry and physical conditions of the atmosphere without expectations of physical consistency. This more flexible approach provides the opportunity to capture conditions that would otherwise be prohibited by self-consistent models, such as combinations of gases not considered under chemical equilibrium. Nonetheless, caution must be exercised in the interpretation of free retrievals as model assumptions may contribute to biased and unphysical atmospheric estimates \citep{welbanks2022terminators}. 

We employed the retrieval framework \texttt{Aurora} \citep{welbanks2021aurora} capable of modeling and interpreting transmission and emission spectra of transiting exoplanets \citep{welbanks2022terminators, bell23_wasp80b}. A detailed description of \texttt{Aurora}'s transmission modeling approach is described in \cite{welbanks2021aurora}. Briefly, the atmospheric model solves radiative transfer for a parallel-plane atmosphere under hydrostatic equilibrium for a transmission geometry. The atmospheric model considers a one-dimensional model atmosphere spanning from $10^{-6}$ bar to 100 bar, divided into 100 layers uniformly spaced in logarithmic pressure space. The vertical temperature structure is parameterized following \cite{madhu2009retrievals}. The model retrieves the reference pressure for an assumed planetary radius of 0.43 $R_J$.

The atmospheric models assume vertically uniform mixing ratios for H$_2$O \citep{Rothman2010}, CH$_4$ \citep{Yurchenko2014}, NH$_3$ \citep{Yurchenko2011}, HCN \citep{Barber2014},  CO \citep{Rothman2010}, and  CO$_2$ \citep{Rothman2010} using independent free parameters for each gas' volume mixing ratio. Our model also considers the effects of H$_2$–-H$_2$ and H$_2$–-He collision induced absorption \citep[CIA;][]{Richard2012}. The presence of clouds and hazes is incorporated via a one-sector parameterization \citep{welbanks2021aurora}, which treats the combined spectroscopic effect of an optically thick cloud deck at a pressure P$_\mathrm{cloud}$ together with hazes following an enhancement to Rayleigh-scattering \citep{Lecavelier2008} in a linear combination with a cloud-free atmosphere \citep{line2016nonuniformclouds}. 

The Bayesian inference is performed using nested sampling \citep{Skilling2004,Feroz2009} via \texttt{PyMultiNest} \citep{Buchner2014} using 500 live points in the sampling and the default parameters for the sampler. Each forward model in the sampling was calculated using line-by-line opacity sampling from 1.0$\mu$m to 1.8$\mu$m using 5000 linearly spaced spectral points for an average spectral resolution of $\sim8500$. In total, the sampling is performed over 17 parameters: 6 molecular gases, 6 for the pressure-temperature structure of the planet, 1 for the reference pressure, and 4 for the presence of inhomogeneous clouds and hazes. 

\section{Results}
\label{sec:results}

\subsection{Joint Broadband Transit, RV, and SED Fitting}
\label{subsec:jointBBfit}

\begin{figure*}[ht!]
    \centering
    \includegraphics[width=\textwidth]{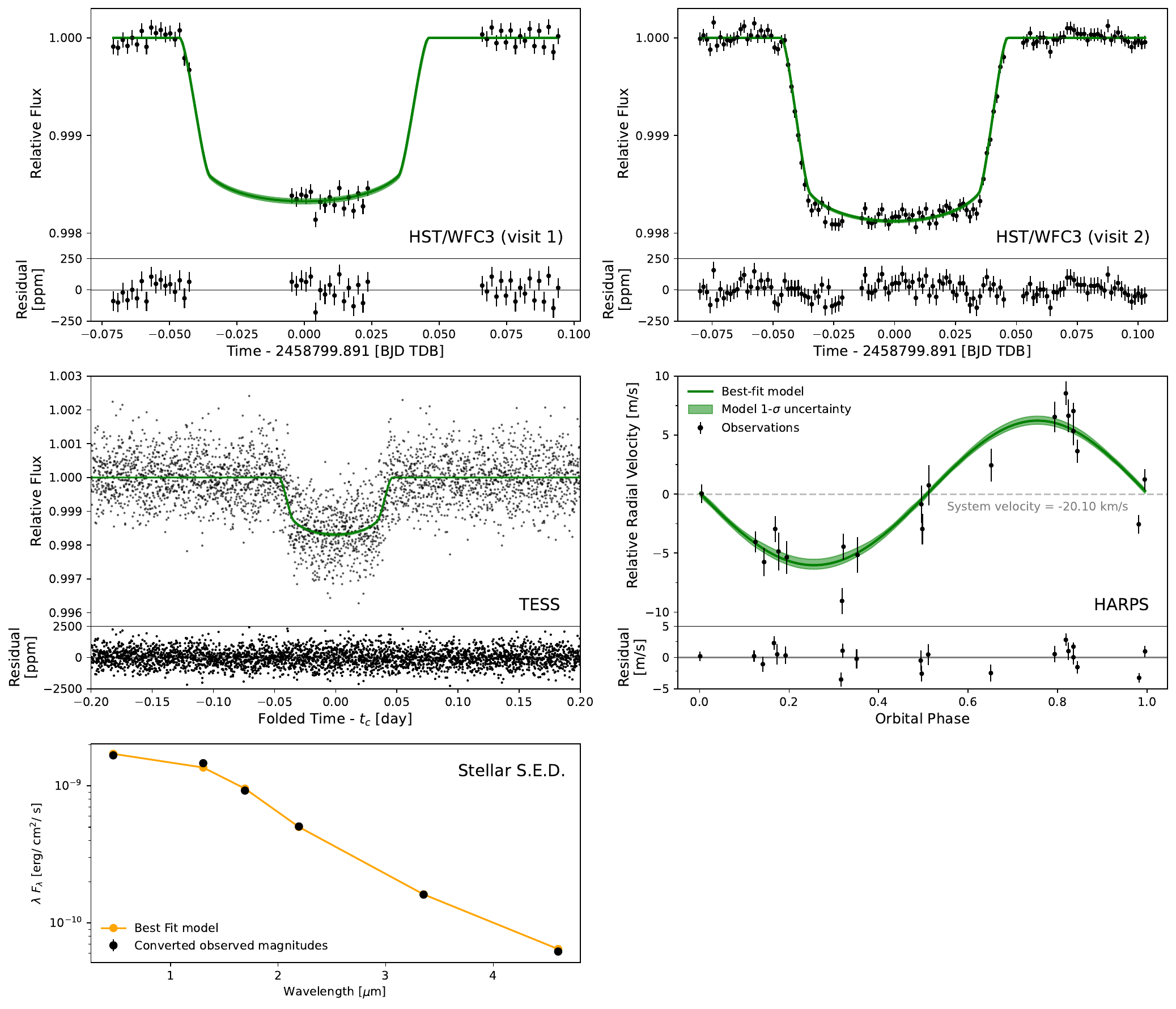}
    \caption{Detrended observed data, best-fit models, and data-model residuals from our simultaneous fit of the broadband transit and radial velocity data. The HST/WFC3 broadband light curves from each visit are shown in the top row, and the 12 TESS light curves, after phase folding them all together, are shown in the middle-left panel. The HARPS RV data from \cite{esposito2019} is shown in the middle-right panel. Also shown in the bottom row are the observed fluxes of the star HD 219666, converted from the magnitudes listed in Table~\ref{tab:magnitudes}, along with our best-fitting stellar SED model. From these simultaneous fits, we precisely constrained the properties of HD~219666~b's orbit.}
    \label{fig:jointfitgallery}
\end{figure*}

Our simultaneous fit of the broadband HST/WFC3, TESS, and RV observations tightly constrain HD~219666~b's orbit. In particular, we refine HD~219666~b's orbital period to a 1-$\sigma$ precision of 0.3~seconds (P = 6.034468 $\pm$ 0.000004). Our period measurement is consistent with that of \cite{hellier2019} based on WASP-South observations, and our semimajor axis and inclination measurements are consistent with \cite{hellier2019} based on the 2018 TESS observations. There has been notable disagreement on these parameters across the literature, discussed further in Section~\ref{subsec:litcomparison}, that led to the several missed transits during our HST programs. Our new constraints on HD~219666~b's orbit are based on the most extensive, precise, and self-consistently fit set of observations to date. Figure~\ref{fig:jointfitgallery} shows the detrended data and best-fit models for each, and Table~\ref{tab:JFvals} gives the best-fit values for our fitting parameters and other useful parameters derived therefrom. We show the full corner plot of the MCMC results in the Appendix, Figure~\ref{apxfix:jfcornerplot}. Our best-fit SED model is also shown in Figure~\ref{fig:jointfitgallery}, and our constraints on the HD~219666's properties are also given in Table~\ref{tab:JFvals}. The stellar temperature and radius that we derive are consistent with those from \cite{esposito2019}, though we infer a slightly higher stellar mass. 

\begin{table*}
\caption{Bayesian prior and parameter posterior values from our simultaneous fit of the HST and TESS transit data, and HARPS RV data of HD 219666 b, as well as our fit of the stellar spectral energy distribution.} 
\centering
\label{tab:JFvals}
\begin{tabular}{lccccl}\toprule 
Parameter & Units & Prior & Best-fit Value \\ \midrule 
\textit{Fitted Parameters} \rule[2pt]{1in}{0.5pt}   & & &  \\
$t_c$                      & Time of conjunction [BJD$_{\text{TDB}}$]   &  None             &  2459083.5107 $\pm$  0.0001 \\
$\log_{10} \left(P\right)$ & log Orbital period [days]                  &  None             &  0.7806389 $\pm$ 2 $\times$ 10$^{-7}$ \\
$\log_{10} \left(a / R_\star \right)$ & log Scaled semi-major axis      &  None             &  1.12 $\pm$ 0.01 \\
$\cos ~i$                  & Cosine of inclination angle                &  None             &  0.062 $\pm$ 0.003   \\
$R_p / R_\star$            & Radius ratio, HST broadband (Visit 1)      &  None             &  0.0415 $\pm$ 0.0004 \\
$R_p / R_\star$            & Radius ratio, HST broadband (Visit 2)      &  None             &  0.0439 $\pm$ 0.0002 \\
$R_p / R_\star$            & Radius ratio, TESS                         &  None             &  0.0421 $\pm$ 0.0006 \\
$K$                        & RV semi-amplitude [m/s]                    &  None             &  6.07 $\pm$ 0.32  \\
$\gamma$                   & System velocity [km/s]                     &  None             &  -20.0975 $\pm$ 0.0002  \\
$\sqrt{e} \cos \left( \omega \right) $ &                                &  None             &   0.17 $\pm$ 0.16  \\
$\sqrt{e} \sin \left( \omega \right) $ &                                &  None             &   -0.0003 $\pm$ 0.1225  \\
$T_{eff}$                  & Stellar effective temperature [K]          &  5527 $\pm$ 65    &  5540 $\pm$ 63 \\
$R_\star$                  & Stellar radius [R$_\odot$]                 &  None             &   1.059 $\pm$ 0.008 \\
$\mu$                      & System parallax [mas]                      &  10.62 $\pm$ 0.01 &  10.62 $\pm$ 0.01 \\
$A_V$                      & V-band extinction [mag]                    &  None             &  0.19 $\pm$ 0.05 \\
$u_{1, TESS}$              & Limb darkening coefficient (TESS)          &  0.39 $\pm$ 0.05  & 0.39 $\pm$  0.04 \\
$u_{2,TESS}$               & Limb darkening coefficient (TESS)          &  0.25 $\pm$ 0.05  & 0.26 $\pm$ 0.05 \\
$u_{1, HST}$               & Limb darkening coefficient (HST)           &  0.18 $\pm$ 0.01  & 0.17 $\pm$ 0.01 \\
$u_{2, HST}$               & Limb darkening coefficient (HST)           &  0.24 $\pm$ 0.01  & 0.21 $\pm$ 0.01 \\
$A_1$                      & HST ``hook" amplitude (Visit 1)            &  None             &  0.00147 $\pm$ 4 $\times$ 10$^{-5}$ \\
$A_2$                      & HST ``hook" amplitude (Visit 2)            &  None             &  0.00020 $\pm$ 3 $\times$ 10$^{-5}$ \\
$\tau_1$                   & HST ``hook" timescale (Visit 1)            &  None             &  -0.0060 $\pm$ 0.0004 \\
$\tau_2$                   & HST ``hook" timescale (Visit 2)            &  None             &  -0.0035 $\pm$ 0.0009 \\
$s_{t,1}$                  & HST ramp slope (Visit 1)                   &  None             &  -0.0022 $\pm$ 0.0002  \\
$s_{t,2}$                  & HST ramp slope (Visit 2)                   &  None             &   -0.0058 $\pm$ 0.0001 \\
$b_1$                      & HST ramp intercept (Visit 1)               &  None             &  1.00071 $\pm$ 3 $\times$ 10$^{-5}$ \\
$b_2$                      & HST ramp intercept (Visit 2)               &  None             &  0.99992 $\pm$ 1 $\times$ 10$^{-5}$ \\
&&&\\
\textit{Derived Parameters} \rule[2pt]{1in}{0.5pt} & & & \\
$P$                        & Orbital period [days]                      & N/a               & 6.034468 $\pm$ 0.000004 \\
$a/R_\star$                & Scaled semi-major axis                     & N/a               & 13.25 $\pm$ 0.46 \\
$i$                        & Inclination angle [deg]                    & N/a               & 86.42 $\pm$ 0.23 \\
$e$                        & Eccentricity                               & N/a               & 0.05 $\pm$ 0.04  \\
$\omega$                   & Argument of Periastron [deg]               & N/a               & 0 $\pm$ 48 \\
$M_p$                      & Planet mass [M$_J$ / M$_\earth$]                 & N/a               & 0.055 $\pm$ 0.003 / 17.6 $\pm$ 1.2 \\
$R_p$                      & Planet radius, TESS [R$_J$ / R$_\earth$]         & N/a               & 0.437 $\pm$ 0.006 / 4.898 $\pm$ 0.06 \\
$R_p$                      & Planet radius, HST Visit 1 [R$_J$ / R$_\earth$]  & N/a               & 0.428 $\pm$ 0.003 / 4.797 $\pm$ 0.03 \\
$R_p$                      & Planet radius, HST Visit 2 [R$_J$ / R$_\earth$]  & N/a               & 0.453 $\pm$ 0.002 / 5.078 $\pm$ 0.02 \\
$M_\star$                  & Stellar mass [M$_\odot$]                   & N/a               & 1.02 $\pm$ 0.1 \\
$T_{eq}$                   & Zero-albedo equilibrium temperature [K]   & N/a                & 1076 $\pm$ 22 
  \\\bottomrule 
\end{tabular}
\tablecomments{Also given are relevant parameters derived therefrom. All priors given are Gaussian. Best-fit values represent the median of their corresponding posterior distributions, and uncertainties are the mean of the posterior's 16th and 84th percentiles.}
\end{table*}

Our fit of the RV observations yields an eccentricity of $e$ = 0.05 $\pm$ 0.04, which is technically nonzero but consistent with zero within 1.25-$\sigma$. \cite{esposito2019} found a similar value of $e$ = 0.07$^{+0.06}_{-0.05}$ from these same data, which is consistent with zero within 1.4-$\sigma$. \cite{esposito2019} note that there is a 35\% probability that, even if HD~219666~b's orbit is circular, white noise in the data could have led their Monte Carlo-based fit to such a nonzero eccentricity. This probability is even larger for our measurement and exceeds the sub-5\% significance threshold recommended by \cite{lucy1971_rvbias}, suggesting that there is not sufficient evidence that HD~219666~b's orbit is actually noncircular. Assuming a Neptune-like tidal quality factor of $Q$ = 10$^4$, using Equation~2 of \cite{jackson2009_tides} we estimate that HD~219666~b's orbit circularization timescale should be $\sim$300~Myr, far shorter than the estimated $\sim$10~Gyr age of the system \citep{esposito2019}. It is therefore unlikely that HD~219666~b's eccentricity is truly nonzero, though future secondary-eclipse observations would be necessary to conclusively determine this. It is worth nothing that other short-period Neptunian-sized planets, such as GJ~3470~b \citep{kosiarek2019_gj3470b}, have been found to have nonzero eccentricities. With this in mind, to account for potential error in HD~219666~b's eccentricity, when deriving the planet's transmission spectrum later on we ran parallel cases fixing the eccentricity to $e$ = 0.05 $\pm$ 0.04 and $e=0$. We found that the spectra did not significantly change between either case, so this uncertainty in $e$ has no impact on our results. For simplicity, then, in the results shown later we assume a circular orbit.

\begin{figure}
    \centering
    \includegraphics[width=\columnwidth]{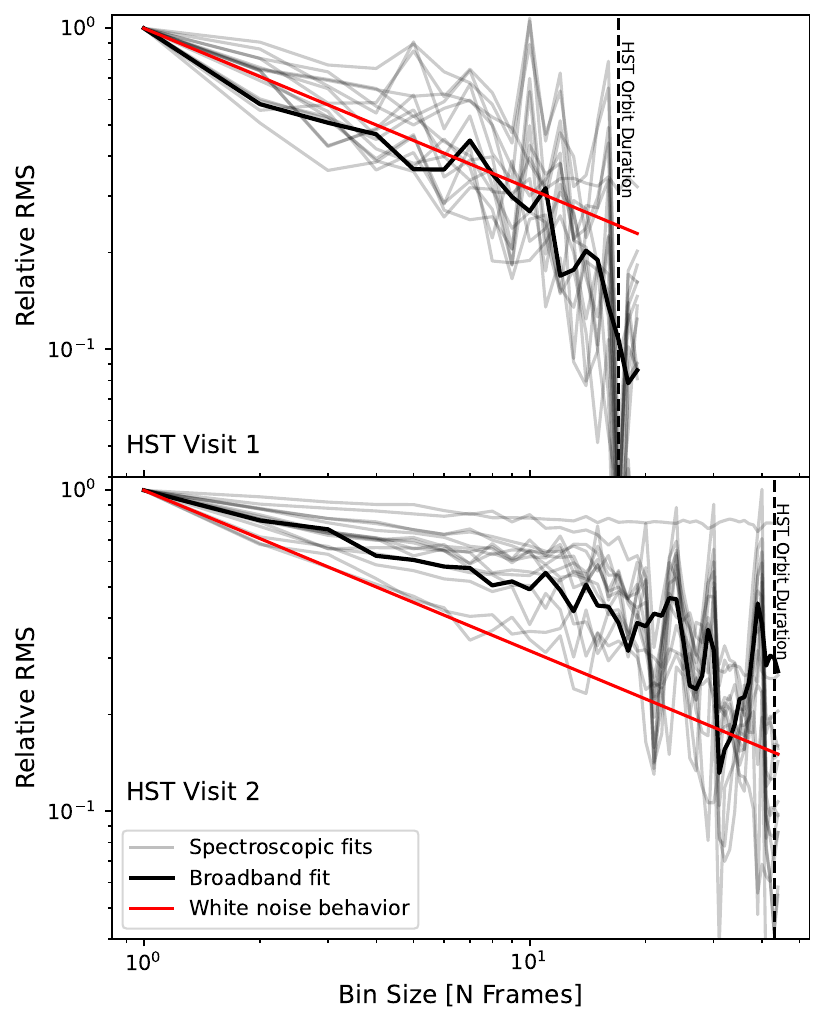}
    \caption{Allan variance plots, showing the RMS of residuals as a function of temporal bin size (N), for each of our fits to the two HST observations presented in this work. The black lines were calculated for our fits to each visit's broadband light curve, and the gray lines for each visit's spectroscopic light curves. The first visit's residuals are consistent with pure white noise whose RMS falls off as $1/\sqrt{N}$. The second visit's RMS values fall off shallower than the white-noise expectation, suggesting excess red noise is present in the data.
    }
    \label{fig:AVplots}
\end{figure}

We find two significantly different values for the HST/WFC3 G141 broadband transit depth from our two visits. For the first visit, in 2019 November, we find a planet-to-star radius ratio of $R_p$/$R_\star$ = 0.0415 $\pm$ 0.0004, or a transit depth of 1722 $\pm$ 33~ppm. For the second visit, in 2020 May, we find a larger value of $R_p$/$R_\star$ = 0.0439 $\pm$ 0.0002, or a transit depth of 1927 $\pm$ 17~ppm. These two values are discrepant at $\sim$5.5-$\sigma$ and, if astrophysical, would mean a 205~ppm variation in transit depth over $\sim$6 months or $\sim$30 orbital periods of HD~219666~b. A systematic origin is likely as well. The residuals from the first HST visit are consistent with pure white noise up to temporal bin sizes comparable to the HST orbit duration, as shown in Figure~\ref{fig:AVplots}. However, those of the second HST visit exhibit evidence for residual red noise. We discuss this offset further in Section~\ref{subsec:starpulse}. Recent work by \cite{edwards24_hstoffsets} has also suggested that the specific functional form of the long-term ramp applied to the HST light curves, in our case a linear function, can affect the measured transit depth. We therefore repeated our fits to the broadband HST light curves testing two additional ramp functions: a quadratic polynomial and a double-exponential ramp \citep[as in][]{deWit18_exponentialhstramp}. The double-exponential ramp did not successfully converge to a solution for either HST visit. The quadratic ramp did converge and yielded an increased transit depth for both visits, and reduced the offset between visits to approximately 110~ppm. However, the uncertainty on each transit depth increased by 2.7$\times$ and 1.8$\times$ for each visit, respectively, and a comparison of the Bayesian Information Criterion preferred the linear ramp in both cases. Therefore, our choice of a linear ramp does not seem to be the sole cause of this offset.

\subsection{Observed Transmission Spectrum of HD 219666 b}
\label{subsec:transmissionspectrum}

\begin{figure*}
    \centering
    \includegraphics[width=\textwidth]{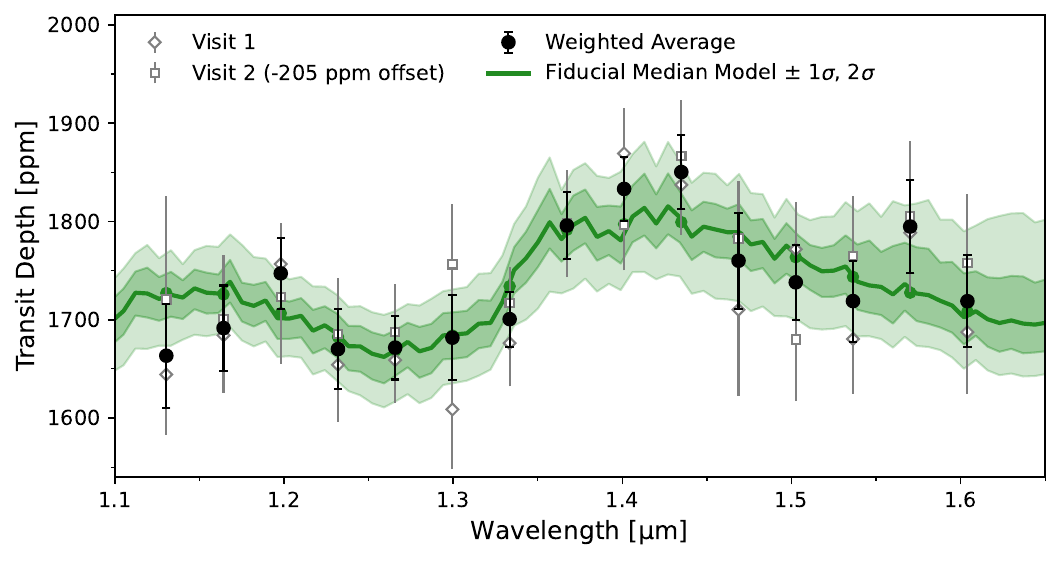}
    \caption{The near-IR transmission spectrum of HD~219666~b observed over two visits with HST/WFC3 G141. The spectra derived from each individual visit are shown as the gray diamond and square points, with a uniform 205~ppm offset applied to all points from the second visit. The error-weighted average of our two visits is shown as the black points, and the median retrieved model spectrum is shown as the green line with its 1 and 2-$\sigma$ uncertainties represented by the green shaded regions. }
    \label{fig:transpec}
\end{figure*}

We separately derived the transmission spectrum from each of our two HST visits, shown as the gray diamond and square points in Figure~\ref{fig:transpec}. Example corner plots for the 1.4~$\mu$m bin from the MCMC runs for each visit are shown in the Appendix, Figures~\ref{apxfig:speccornerplot1} and \ref{apxfig:speccornerplot2}. As with the broadband fits, we found an offset in the absolute spectroscopic transit depths between the two visits. However, we found that the shape and relative feature amplitudes of each spectrum were very consistent within our achieved uncertainties, which is clear when we align the spectra by vertically offsetting Visit 2's points downward by 205~ppm (the difference between the broadband transit depths). We find no evidence that this offset is chromatic, so this offset does not affect our ability to model and interpret HD~219666~b's transmission spectrum. We combined these two spectra by applying this uniform 205~ppm downward offset to Visit 2's spectrum, then calculating the uncertainty-weighted mean of the two spectra, accounting for these weights when propagating the transit depth uncertainties. We show this combined spectrum as the black points in Figure~\ref{fig:transpec}, and tabulate all transit depth values in Table~\ref{tab:transitdepthtable}. 

\begin{table}[h!]
    \centering
    \caption{Spectroscopic transit depths of HD~219666~b as measured by HST/WFC G141. }
    \begin{tabular}{c|c|c|c}
        Wavelength [$\mathrm{\mu}$m] & Visit 1 & Visit 2 & Combined \\ \hline 
        1.130 $\pm$ 0.017 & 1644 $\pm$ 61 & 1926 $\pm$ 106 & 1663 $\pm$ 53 \\
        1.164 $\pm$ 0.017 & 1684 $\pm$ 58 & 1906 $\pm$ 65 & 1692 $\pm$ 43 \\
        1.198 $\pm$ 0.017 & 1757 $\pm$ 42 & 1928 $\pm$ 68 & 1747 $\pm$ 36 \\
        1.232 $\pm$ 0.017 & 1654 $\pm$ 58 & 1891 $\pm$ 57 & 1670 $\pm$ 41 \\
        1.266 $\pm$ 0.017 & 1659 $\pm$ 43 & 1893 $\pm$ 49 & 1672 $\pm$ 32 \\
        1.299 $\pm$ 0.017 & 1609 $\pm$ 61 & 1961 $\pm$ 61 & 1682 $\pm$ 43 \\
        1.333 $\pm$ 0.017 & 1676 $\pm$ 44 & 1922 $\pm$ 36 & 1701 $\pm$ 28 \\
        1.367 $\pm$ 0.017 & 1794 $\pm$ 43 & 2003 $\pm$ 55 & 1796 $\pm$ 34 \\
        1.401 $\pm$ 0.017 & 1869 $\pm$ 46 & 2002 $\pm$ 46 & 1833 $\pm$ 33 \\
        1.435 $\pm$ 0.017 & 1837 $\pm$ 51 & 2072 $\pm$ 57 & 1850 $\pm$ 38 \\
        1.469 $\pm$ 0.017 & 1710 $\pm$ 87 & 1988 $\pm$ 59 & 1760 $\pm$ 49 \\
        1.503 $\pm$ 0.017 & 1772 $\pm$ 48 & 1885 $\pm$ 63 & 1738 $\pm$ 38 \\
        1.536 $\pm$ 0.017 & 1680 $\pm$ 56 & 1970 $\pm$ 61 & 1719 $\pm$ 41 \\
        1.570 $\pm$ 0.017 & 1788 $\pm$ 60 & 2010 $\pm$ 77 & 1795 $\pm$ 47 \\
        1.604 $\pm$ 0.017 & 1687 $\pm$ 63 & 1963 $\pm$ 70 & 1719 $\pm$ 47
    \end{tabular}
    \tablecomments{All transit depths are given in parts per million (ppm). The combined spectrum is the uncertainty-weighted average of the the two individual visits once Visit~2 is offset by 205~ppm.}
    \label{tab:transitdepthtable}
\end{table}

To infer the properties of HD~219666~b's atmosphere from our observed transmission spectrum, we performed a free retrieval on the combined spectrum using \texttt{Aurora}, described in Section~\ref{sec:methods_retrieval}. We show the resulting median retrieved model as the green line in Figure~\ref{fig:transpec}, with the shaded green regions representing the 1- and 2-$\sigma$ model uncertainty, and provide this retrieval's full posterior distribution in the Appendix,  Figure~\ref{apxfig:fiducialretrievalcornerplot}.

We detect the presence of water in HD~219666~b's atmosphere at 2.9-$\sigma$ significance, and infer a volume mixing ratio of $\log_{10}$X$_{\text{H}_2\text{O}}=-1.43 ^{+ 0.53 }_{- 1.29 }$. The detection of H$_2$O refers to a preference for our fiducial model (Bayesian evidence of $\log \mathcal{Z}=125.56$) relative  to a model without H$_2$O ($\log \mathcal{Z}=122.65$) in it. This comparison of nested models using their respective Bayesian evidence  is commonplace in atmospheric studies as a result of the implementation of nested sampling algorithms in atmospheric retrievals \citep[e.g.,][]{Benneke2013} and mappings for their associated Bayes factor \citep[see, e.g., equation 17 in ][]{welbanks2021aurora}. However, the exclusive use of single value statistics to infer the presence of absorbers in a spectrum may result in flawed and conflicting inferences \citep[see, e.g.,][for a discussion]{Welbanks2023}.

Further comparisons of nested models suggest that the presence of methane is not preferred based on the evidence for the model without methane ($\log \mathcal{Z}=125.72$) and its unconstrained abundance. Likewise, the abundance of NH$_3$ is not constrained. A marginal model preference for HCN at 1.6-$\sigma$ is present, although the gas' abundance is unconstrained. Overall, besides suggestions of H$_2$O absorption, other parameters in the model remain largely unconstrained, including the cloud and haze parameters in the model. The retrieved terminator temperature at 100~mbar is weakly constrained to 1105$^{+277}_{-292}$~K and consistent with our derived equilibrium temperature of 1076 $\pm$ 22~K based on our fit to HD~219666~b's orbit (Table~\ref{tab:JFvals}). Our use of a nonisothermal parametric temperature-structure prescription is not significantly preferred over an isothermal vertical temperature structure and the inferred abundances are consistent within their uncertainties regardless of this modeling choice (e.g., $\log_{10}$X$_{\text{H}_2\text{O}}=-1.57 ^{+ 0.66 }_{- 2.05 }$, T$_\text{isotherm}=900 ^{+ 224 }_{- 225 }$~K in the isothermal case). Our choice of non-isothermal profiles is motivated by works showing that assuming isothermal profiles can lead to significant biases in retrieved temperatures and abundances \citep[e.g.][]{Welbanks2019a, welbanks2022terminators}.

Altogether, these results suggest that HD~219666~b has a water-rich, methane-depleted atmosphere. Our inferred water abundance is  consistent with that of \cite{edwards2023_popstudy}, who independently measured a visually similar transmission spectrum of just Visit 1 and determined a value of $\log_{10}$X$_{\text{H}_2\text{O}}=-1.54 ^{+ 0.37 }_{- 1.08 }$ therefrom.
 
The apparent lack of methane in HD~219666~b's atmosphere is reminiscent of similar nondetections with HST/WFC3 on K2-18~b \citep{benneke19_k218b}, GJ~3470~b \citep{benneke2019_gj3470}, TOI-270~d \cite{mikalevans23_toi270d}, and even the sub-Saturn WASP-80~b \citep{wong22_wasp80b}. Even though HD~219666~b is hotter than these other planets, methane should be significantly abundant in its atmosphere under chemical equilibrium, with a volume mixing ratio potentially ranging up to $\sim$10$^{-3}$ depending on its carbon-to-oxygen ratio \citep[e.g.,][]{moses2013_eqabundances}, which would have been detectable from our analysis. This points to the action of disequilibrium chemistry in HD~219666~b's atmosphere, driven potentially by strong vertical mixing or even tidal heating given HD~219666~b's possible nonzero eccentricity. 

\section{Discussion}
\label{sec:discussion}

\subsection{Where is the methane?}
\label{subsec:alternateatmo}

Our fiducial retrieval suggests that HD~219666~b has a water-rich atmosphere that lacks or is significantly depleted in methane. However, we must interpret this in the context of recent JWST observations of similar planets that have detected methane despite HST/WFC3 nondetections. Some JWST analyses, on K2-18~b for example \citep{madhusudhan2023_k218b}, even prefer a methane-rich and water-free atmosphere despite previous WFC3 observations having found the opposite. With these recent results in mind, we here explore a hypothetical scenario where HD~219666~b's transmission spectrum is instead shaped by methane absorption.

\begin{figure*}
    \centering
    \includegraphics[width=\textwidth]{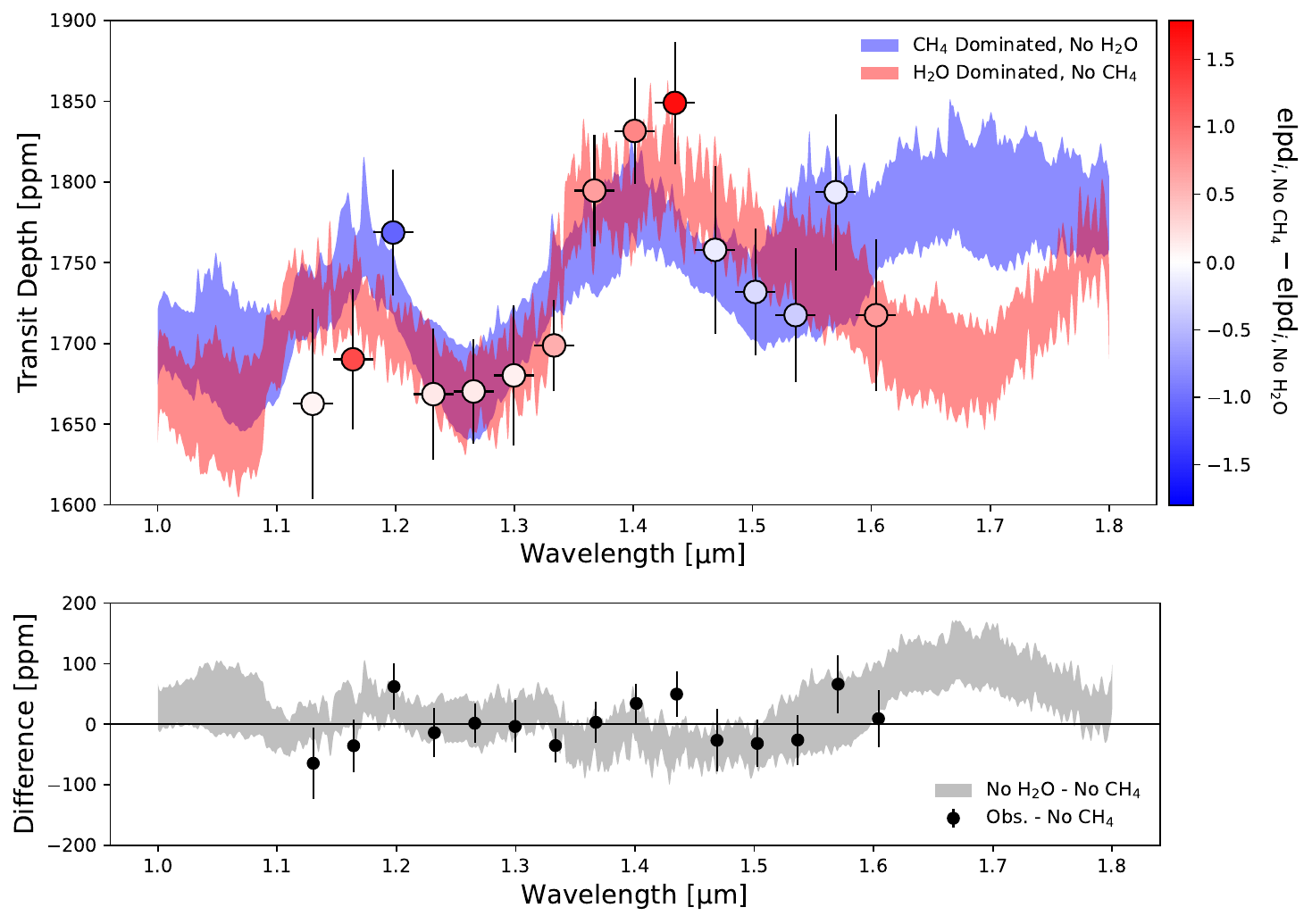}
    \caption{Top: we compare the 1$\sigma$ confidence intervals from a retrieved model spectrum dominated by water absorption and excluding methane (red) to an alternate scenario where the atmosphere has no water and is dominated by methane absorption (blue). The points with error bars are the observed transmission spectrum, as in Figure~\ref{fig:transpec}. We performed a leave-one-out cross-validation analysis (LOO-CV) on the observed spectrum to evaluate each point's relative evidence for the fiducial model. The resulting $\Delta$elpd$_i$ values are presented as the coloring of each data point. For reference, the bluest point has $\Delta$elpd$_i$ = -1.1 and the reddest point has a $\Delta$elpd$_i$ = 1.7. Bottom: We compare the data-model residuals relative to the water-dominated/methane-free model, shown by the black points, to the difference between the methane-free and water-free models, shown as the gray shaded region.
    }
    \label{fig:loocvresults}
\end{figure*}

To explore this water-free scenario, we perform an atmospheric retrieval considering a model atmosphere that does not have water absorption. Fortunately, we already have such a model for free as part of the nested models required for the detection significance calculations in Section~\ref{subsec:transmissionspectrum}. As previously mentioned, the detection significances of each molecule in our fiducial retrieval are calculated by comparing the results of separate retrievals with and without that specific molecule. This means that in calculating the detection significance of water, we had run an additional retrieval where the model atmosphere is forced to exclude water, so that it attempts to explain the observed spectrum with a combination of CH$_4$, HCN, and NH$_3$. In this case, as expected, methane became the dominant contribution to the spectrum, with a median abundance of $\log_{10}$X$_{\text{CH}_4}= -2.35 ^{+ 1.07 }_{- 1.86 }$. As in the fiducial model, HCN and NH$_3$ did not contribute as significantly, so we posit that this water-free retrieval's result represents a hypothetical scenario where methane absorption is solely responsible for the spectral feature seen in our HST observations. This methane-rich retrieved model is shown as the blue model in Figure~\ref{fig:loocvresults}, where the bounds of the shaded region represent the upper and lower 1$\sigma$ uncertainties. For comparison, shown in red is an analogous model excluding methane, which is similar to our fiducial model. We show the posterior distributions for the water-free retrieval in the Appendix, Figure~\ref{apxfig:noH2Oretrievalcornerplot}.

By definition, this model without H$_2$O is disfavored over the fiducial model at the $\sim$3-$\sigma$ level. While computing reduced $\chi^2$ values for these models is ill-defined since the model has more free parameters than there are data, comparisons of $\chi^2$/N$_{\text{data}}$ support the fact that the fiducial model with H$_2$O included ($\chi^2$/N$_{\text{data}}$=0.65) is a better fit than the nested model without H$_2$O (($\chi^2$/N$_{\text{data}}$=1.09). Further, the model that aims to fit these spectroscopic features without H$_2$O absorption is likely unphysical and in conflict with theoretical expectations \citep[e.g.][]{moses2013_eqabundances} and other observational findings \citep[e.g.][]{bell23_wasp80b} that water indeed exists in significant abundance in exoplanetary atmospheres, even when methane is also present.

To explore this scenario further, we calculate the out-of-sample predictive accuracy of both models, with H$_2$O/without CH$_4$ and without H$_2$O/with CH$_4$, at the resolution of individual data points. Our approach follows the methods of \citet{Welbanks2023} for exoplanetary atmospheric analysis and builds on the increasing adoption of leave-one-out cross-validation (LOO-CV) in astronomical studies to enable interpretable model criticism \citep[e.g.,][]{McGill2023, Nixon2024, Challener2023}. Briefly, we calculate the expected log posterior predictive density (elpd$_\text{LOO}$), which allows us to score each data point on how well it is predicted by a given model conditioned on the rest of the data. Comparing the scores of each model (i.e., calculating the $\Delta$elpd$_\text{LOO}$ scores) allows us to interpret which of the two models is better at explaining each data point.

Figure~\ref{fig:loocvresults} shows the resulting LOO-CV analysis for the two models of interest and the 1-$\sigma$ confidence intervals from the retrieval analysis. The magnitude of the score difference (intensity in the color) suggests a better performance by one of the models, while the sign of this difference (positive/red: model with water and without methane; negative/blue: model without water and with methane) indicates which of the models performs better. While the performance of both models is comparable for most data points, three data points indicate a significant improvement in the performance of one of the models. For instance, the sixth point from the left (at $\sim$1.42~$\mu$m) strongly prefers water with a $\Delta$elpd$_i$ = 1.7. The two points to its left also favor water, though less strongly ($\Delta$elpd$_i$ = 0.7 and 0.8, respectively), altogether suggesting that the amplitude and shape of this feature at 1.4~$\mu$m is best explained by water absorption, in agreement with expectations. Similarly, the second point from the left (at $\sim$1.15~$\mu$m) suggests a better performance by the model including water absorption, with a $\Delta$elpd$_i$ = 1.3. On the other hand, only one point, third from the left ($\sim1.2\mu$m), is significantly better explained by the model with methane and no water with a $\Delta$elpd$_i$ = -1.1. For reference, the other points suggesting a better performance by the model with methane and no water have significantly smaller magnitudes in their $\Delta$elpd$_\text{LOO}$ scores of no more than $\sim30\%$ that of the point at $\sim1.2\mu$m. The increased performance of the model with methane is likely due to the slightly higher amplitude of methane's 1.15~$\mu$m absorption feature over that of water. We further contextualize the relative importance of the point at $\sim1.2\mu$m in driving our interpretation of this planetary spectrum by calculating the residuals in the data relative to the methane-free model. The results shown in the bottom panel of Figure~\ref{fig:loocvresults} suggest that the residual in this one point of interest is comparable to the overall scatter of residuals across the entire spectrum.

All together, the presence of water in HD~219666~b's atmosphere is strongly preferred via the Bayesian evidence. Nevertheless, this exercise shows that an alternative, K2-18~b-esque methane-rich atmosphere \textit{can} adequately fit the data, despite being unlikely. As shown by the models in Figure~\ref{fig:loocvresults} and discussed previously in the literature \citep[e.g.][]{blain2021k218bmodels, bezard2022methanevwater}, water and methane both have nearly identical absorption spectra within the G141 bandpass, which contributes to this potential ambiguity. Within this bandpass, there are only two regions where our water-free and methane-free model spectra slightly differ: around 1.19~$\mathrm{\mu}$m, where methane has a narrow peak that water does not, and from 1.4 - 1.5~$\mathrm{\mu}$m, where the methane bump's tail is lower. As shown in the lower panel of Figure~\ref{fig:loocvresults} though, these differences are only on the order of tens of parts per million and are comparable to or smaller than our transit depth uncertainties. Just redward of the G141 bandpass, from ~1.6 - 1.8~$\mu$m, the difference between the water and methane models becomes much greater ($>$100~ppm) where methane has a strong absorption feature that is not overlapped by a water feature. There is also a strong separated methane feature around 3.3~$\mu$m \citep[e.g.]{bell23_wasp80b}. These features could easily be probed by a single JWST observation using NIRISS or NIRCam. Therefore, follow-up observation with JWST would be able to more precisely determine HD~219666~b's atmospheric composition and unveil the true relative abundances of water and methane.

\subsection{Implications of the water--methane ambiguity from HST/WFC3}
\label{subsec:discussion_implications}

So far, five Neptunian-sized worlds have been characterized in transmission using both HST/WFC3 and JWST: K2-18~b \citep{benneke19_k218b, tsiaras2019_k218b, madhusudhan2023_k218b, wogan2024_k218b}; GJ~3470~b \citep{benneke2019_gj3470, beatty24_gj3470b}; TOI-270~d \citep{mikalevans23_toi270d, benneke24_toi270d}; GJ~1214~b \citep{berta2012_gj1214b, kreidberg2014_gj1214b, kempton2023_gj1214b, schlawin2024_gj1214b, ohno2024_gj1214b}, and LTT-9779~b \citep{edwards2023_ltt9779b, radica2024_ltt9779b}. The chemical composition of GJ~1214~b's and LTT-9779~b's atmosphere have been difficult to characterize due to extreme aerosols or metallicity that mute their transmission spectrum and are tangential to this discussion, so we do not consider them further here. Also, LTT-9779~b is sufficiently hot at an equilibrium temperature near 2000~K \citep{jenkins2020_ltt9779b} that methane should not be present in its atmosphere in any case. For the other three planets, the corresponding HST/WFC3 analyses also did not detect methane \citep{benneke2019_gj3470, tsiaras2019_k218b, benneke19_k218b, mikalevans23_toi270d} but the follow-up JWST analyses showed that methane and water actually coexist in their atmospheres \citep{wogan2024_k218b, beatty24_gj3470b, benneke24_toi270d}. Follow-up JWST observations are needed to determine whether the same is true of HD~219666~b. Generally, these analyses are beginning to suggest that HST/WFC3 may have systematically underestimated the prevalence of methane in exoplanetary atmospheres, at least for planets at the temperatures where methane is expected to be significantly abundant. At the very least, these analyses underscore how difficult it is to precisely determine the atmospheric composition of an exoplanet without observations spanning a wide range of wavelengths. Since much of the community's current population-level understanding of exoplanetary atmospheres, and how present-day atmospheres link to planet formation, are based on water abundance measurements from HST/WFC3, this raises several important questions. 

One question raised is whether any reported detections of water with HST/WFC3 could instead be masquerading methane, as posed for K2-18~b by \cite{madhusudhan2023_k218b}. Fortunately, this does not seem to be a concern based on the JWST results to date. None of the water detections on the ``missing methane" planets (e.g., those discussed above, as well as WASP-80~b \citep{wong22_wasp80b, bell23_wasp80b}) have been indisputably overruled, including K2-18~b \citep{wogan2024_k218b}. Rather, JWST is revealing that water and methane coexist in these atmospheres, with methane generally present at depleted but still significant abundances. 

Even if methane is not fully masquerading as water in HST/WFC3 observations, the lack of detections means that HST/WFC3 analyses likely neglect, or at least underestimate, its contribution to the planet's atmosphere. It follows that the water abundances inferred from HST/WFC3 analyses of planets on which methane is also present may be inaccurate as well. A second question raised, then, is what impact, if any, this has on previously determined population trends. For example, one of the most important trends in exoplanetary science is the mass-metallicity relation \citep[see e.g.,][]{welbanks19_massmetallicity, sun24_massmetallicity}, which provides a link between a planet's formation and its present-day atmospheric properties. The Solar System giant planets' atmospheric metallicities, as determined from their atmospheric \textit{methane} abundances, follow a tight linear correlation with their mass. Unlike these Solar System measurements, this mass-metallicity relation for exoplanets has had to be inferred through atmospheric water abundances. Through a uniform suite of retrievals on literature HST/WFC3 data, \cite{welbanks19_massmetallicity} found that the mass-metallicity relation of giant exoplanets--most of which are also too hot for methane to be present--is similar to that of the Solar System giant planets. However, against expectations, \cite{welbanks19_massmetallicity} found that the metallicities of lower-mass Neptunian-sized exoplanets--most of which are at temperatures where methane is expected to be present--fell significantly below the Solar System trend. However, if HST-only analyses have indeed been systematically misestimating the atmospheric water abundance for these planets, and thus the overall atmospheric metallicity, then this discrepancy may not be as extreme as previously thought. 

\begin{figure}
    \centering
    \includegraphics[width=\columnwidth]{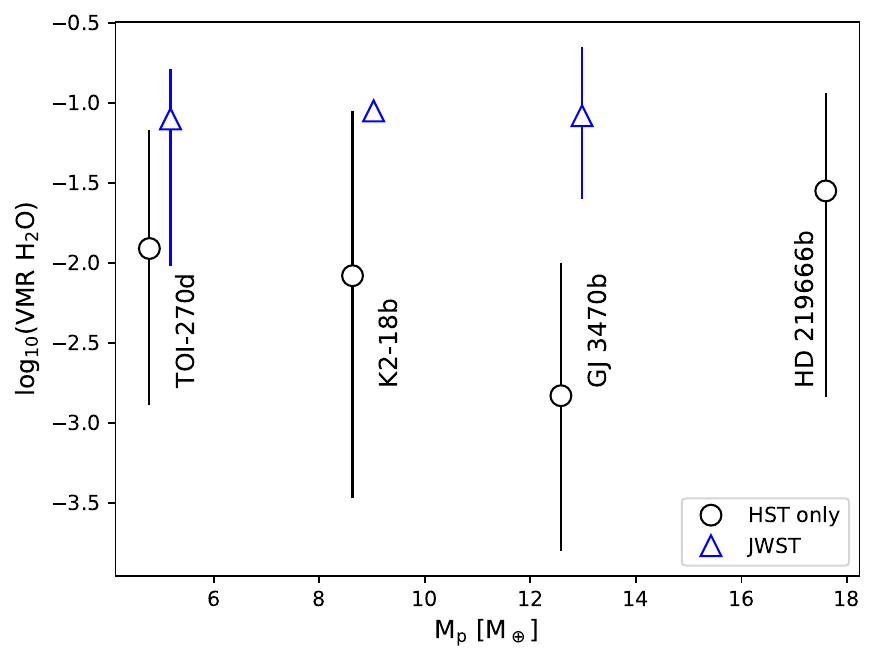}
    \caption{Atmospheric volume mixing ratios of water as inferred from retrievals on HST/WFC3 data (black circles) vs. JWST data (blue triangles) in the literature for planets in the $<$5~R$_\earth$ regime. Note that in this speculation we do not carefully consider exactly which retrieval method or JWST instrument is used. In each case (except our target HD~219666~b, which has not been followed up by JWST), the HST-only analysis underestimated the abundance of water relative to the JWST analysis. 
    Values are taken from \cite{mikalevans23_toi270d} and \cite{benneke24_toi270d} for TOI-270d, \cite{benneke19_k218b} and \cite{wogan2024_k218b} for K2-18b, and \cite{benneke2019_gj3470} and \cite{beatty24_gj3470b} for GJ 3470b. }
    \label{fig:abundances_HSTvJWST}
\end{figure}

To further examine this speculation, we again consider K2-18~b, GJ~3470~b, and TOI-270~d. All three planets are at temperatures where methane is expected in the atmosphere under equilibrium chemistry and, as mentioned, have recently had methane detected by JWST when it was not detected by HST/WFC3. Correspondingly, their inferred water abundances have also changed. Figure~\ref{fig:abundances_HSTvJWST} shows the relative abundances inferred by each instrument for each planet, and we also show our result for HD~219666~b for reference. In all three cases the water abundance inferred by JWST is actually higher than that from HST/WFC3 alone. This is counterintuitive to our speculation that methane has been underestimated, which would suggest that the JWST-based water abundances should instead be lower, highlighting the complexity of this problem, which we discuss further below. We assume that the JWST-based abundances are closer to the truth than the WFC3-only abundances, since JWST probes a wider wavelength coverage (i.e., more absorption features of each molecule) at higher spectrophotometric precision, and because of improvements in modeling techniques over time. With more metals (water and methane) in the atmosphere, it follows that atmospheric metallicities in general should be higher than determined by HST/WFC3. As a result, the lower-mass exoplanets at the relevant temperatures in \cite{welbanks19_massmetallicity}'s mass-metallicity relation may truly be at higher metallicities, and thus more consistent with the Solar System trend, than previously thought. Further work is necessary to confirm this speculation.

As mentioned, the fact that the JWST-based water abundances are each higher than those from HST/WFC3 alone is counterintuitive to the idea that methane may be responsible for some of the absorption in the HST/WFC3 bandpass. In this case, we would expect that the JWST-based abundances should instead be lower. This importantly highlights that the precise determination of molecular abundances is a complex endeavor, and that the underestimation of water and methane are not the only factors in the above discussion. For example, there are important methodological differences between teams that likely account for part of the changes in inferred abundances. Similarly, \cite{lueber2024_jwstretrievals} show that even within a single self-consistent analysis, the abundance that one retrieves can vary depending on the exact model complexity. Also, these JWST measurements may slightly change as more data are taken, and as instrument systematics become better understood. These avenues are already being partially explored, with Welbanks et al. (in prep.) systematically evaluating the differences in retrieval frameworks used by the community, and several ongoing JWST programs following up many sub-Neptune to Neptune-sized planets (e.g., GTO 1185, \cite{jwstprop_1185}; GO 3557 \cite{jwstprop_3557}; GO 4105, \cite{jwstprop_4105}; among others). These new insights and data, coupled with a new, uniform analysis of all of the HST/WFC3 and JWST data together, will shed better light into this issue, and help set straight our understanding of exoplanetary atmospheres. 

\subsection{HD~219666~b within the cloudy-to-clear Neptune landscape}
\label{subsec:cloudtrends}

Besides the ``missing methane" problem, another important open question regards the properties of clouds and hazes in Neptunian-sized planets. Some Neptunian-sized planets like GJ~1214~b \citep{kreidberg2014_gj1214b, kempton2023_gj1214b}, GJ~436~b \citep{knutson2014_gj436b}, and HD~97658~b \citep{knutson2014_hd97658, guo2020_hd97658b} appear to be completely enshrouded by aerosols, while others like GJ~3470~b \citep{beatty24_gj3470b} are not. The reason for these significant differences in cloudiness/haziness and the overall properties of the aerosols on each planet are still not well constrained. 

Several literature works have attempted to shed light into the nature of these clouds/hazes by leveraging population trends, typically using the amplitude of the 1.4~$\mu$m feature in HST/WFC3 observations as a probe of the planet's cloudiness \citep{fu17_hsttrends, crossfield17_hsttrends, edwards2023_popstudy, brande24_clouds}. It is typically assumed that a larger 1.4~$\mu$m feature, relative to the planet's atmospheric scale height, indicates a generally clearer atmosphere. These works identified potential trends between the amplitude of this feature and the planet's equilibrium temperature, but have generally found that the exact relationship is complex and likely involves a number of additional factors. For instance, \cite{crossfield17_hsttrends} and \cite{fu17_hsttrends} initially found a tentative trend of increasing feature strength with increasing temperature, starting at approximately 500~K, for Neptunian-sized planets. When adding additional planets to the sample of \cite{crossfield17_hsttrends}, \cite{edwards2023_ltt9779b} found that $\sim$500~K may be a local minimum, and that feature sizes may again increase as the temperature decreases below this point. \cite{brande24_clouds} found the same general trend in their independent analysis. As the second hottest Neptunian-sized world characterized to date, and having a strong 1.4~$\mu$m feature, HD~219666~b offers a useful additional data point to these analyses.

Following the method of \cite{stevenson2016_quantclouds}, we quantify the amplitude $A_H$ of HD~219666~b's 1.4~$\mu$m feature as 
\begin{equation}
    A_H = \left( \delta_{1.4} - \delta_{1.25} \right) \frac{R_\star^2}{2 H R_p}, \label{eqn:wateramplitude}
\end{equation}
where $\delta_\lambda$ is the observed transit depth at $\lambda$~$\mu$m and $H$ is the atmospheric scale height ($H = k_B T_{eq} / \left( \mu g \right)$) calculated assuming some mean molecular weight. Using the planetary radius from our HST Visit 1 and the combined transmission spectrum, we calculate an amplitude of $A_H$ = 4.77 when assuming $\mu$=3.80 as done by \cite{fu17_hsttrends}, and $A_H$ = 3.83 when assuming $\mu$=3.05 as done by \cite{brande24_clouds}. In either case, to the best of our knowledge, this is the highest-amplitude 1.4~$\mu$m feature of any planet in this regime. This suggests that HD~219666~b likely has a clear atmosphere similar to GJ~3470~b.

To further understand HD~219666~b in context, we place it among the trends inferred by \cite{fu17_hsttrends}, \cite{crossfield17_hsttrends}, \cite{edwards2023_popstudy} and \cite{brande24_clouds}. Note that we are specifically only referring to the subfocus or subsamples of Neptunian-sized exoplanets in these works. With $A_H$ $\geq$ 3.83 at $T_{eq}$ = 1076~K, HD~219666~b lies significantly above the linear $A_H$ versus $T_{eq}$ trend found by \cite{fu17_hsttrends} (see their Figure~2) from a sample of 34 planets, though is similar to the planet XO-1~b (A$_H$ = 3.33 with $\mu$=3.8, $T_{eq}$=1196~K) which is more Jupiter-sized. \cite{crossfield17_hsttrends} derived a similar trend to \cite{fu17_hsttrends}, though with a smaller sample size of only 6 planets, and again HD~219666~b lies significantly above their predicted $A_H$ vs. $T_{eq}$ trend (see their Figure~3). \cite{edwards2023_popstudy} and \cite{brande24_clouds} inferred a potentially parabolic relationship between $A_H$ vs. $T_{eq}$, with $A_H$ steadily increasing with increasing $T_{eq}$ above $\sim$600~K. \cite{edwards2023_popstudy} (see their Figure 10) did not extend their trend fully 1000~K but, extrapolating by eye, HD~219666~b lies above it. HD~219666~b instead falls below the $A_H$ value predicted by \cite{brande24_clouds} (of $A_H \sim$ 6) at 1076~K (see their Figure~3). \cite{edwards2023_popstudy} predicted the approximate feature sizes when clouds exist at specific base pressures, although they assume a range of surface gravities slightly higher than that of HD~219666~b. Nevertheless, HD~219666~b would fall near their expected region with a cloud base at $10^{3}$~Pa. \cite{brande24_clouds} similarly predicted regions in the $A_H$ vs. $T_{eq}$ space where planets should fall given certain assumptions about the cloud sedimentation or haze particle size (see their Figure~3). HD~219666~b would lie in the overlap of all potential cloud sedimentation strengths including the clear case, and lies within the clear atmosphere region of the haze models. All together, these suggest that HD~219666~b's atmosphere is not significantly enshrouded by clouds or hazes, making it a favorable candidate for follow-up with JWST.

\subsection{Possible Origins of our transit depth offset}
\label{subsec:starpulse}

As mentioned in Section~\ref{sec:results}, we find a significant ($>$5$\sigma$) discrepancy between the R$_p$/R$_\star$ values from our two HST/WFC3 G141 visits. The value measured during the second visit is larger by a factor of 1.06x, or 6\%, and corresponds to an increase in transit depth of 205~ppm. If this change in transit depth is solely astrophysical, it would require either the planetary radius or stellar radius, or some combination of the two, to change by 6\% in the 175~days between the visits. 

Percent-level variability in planetary eclipse depths and phase curve properties at optical wavelengths have been claimed previously \citep[e.g.][]{armstrong2016_variabilityHATP7b, jackson2019_variabilityKEPLER76b, changeat2024_variabilityWASP121b}, though skepticism has been raised over whether they are truly linked to the planet's atmosphere \citep{bell2019_variabilityWASP12b, vonessen2019_variabilityWASP12b, lally2022_variabilityHATP7b}. Regardless, to first order, this variability stems from the planet's emission map, and should not significantly affect the planet's radius. Variability in clouds or hazes in the planet's atmosphere could affect the observed radius, but would also chromatically affect the shape and relative amplitude of features of the transmission spectrum rather than uniformly offsetting it as we observe. Tidal interactions with the host star are generally believed to inflate the radii of close-in exoplanets \citep[e.g.,][]{ibgui2009_tidalinflation}, but should not drive such drastic change on this short timescale. Therefore, we find no plausible planetary explanation for the observed offset in transit depth. Radius pulsations and variations are typical for evolved stars. HD~219666 is a $\sim$10~Gyr G-type star \citep{esposito2019}, so it is likely nearing the end stage of its main-sequence lifetime. To explain the increase in transit depth between our HST visits, HD~219666 would have needed to decrease in radius by 6\% over 175~days. Assuming for simplicity that the star is a blackbody emitter, such a radius change would be accompanied by a $\sim$3\% increase in temperature, $\sim$0.6\% increase in luminosity, and a $\sim$5.7 ppt relative increase in magnitude. These two HST visits occurred between the 2018 and 2020 TESS visits, 110~days and 285~days after the first 2018 transit observed by TESS, respectively. If we combine all three TESS visits and normalize each visit's TESS PDCSAP flux by the combined median, rather than normalizing visit-wise, there is an apparent long-term variation in the normalized PDCSAP flux consistent with a 5.7~ppt flux increase between the two HST visits. However, the PDCSAP flux level measured for constant emitters has been found to vary between visits, even in consecutive TESS sectors, due purely to instrumental reasons \citep{nandakumar2022_tesseval}, so this apparent trend is likely just a coincidence. Also, a purely radial pulsation of 6\% on such a relatively short timescale is likely unphysically large for a star like HD~219666. Therefore, we find this stellar pulsation explanation unlikely as well.

Finally, we consider the transit light source effect. Unocculted stellar heterogeneities are known to induce offsets to the observed transmission spectrum between different epochs, or relative to a heterogeneity-free case \citep{rackham18_tlseMtype, rackham19_tlseSunlike}. However, according to the models of \cite{rackham19_tlseSunlike} for a G-type star, the transit light source effect should only bias the observed transit depth by less than 1$\%$, whether it is due only to spots or to both spots and faculae. This is far smaller than the offset we observe. Further, the offsets due to this effect are chromatic and strongly sloped within the G141 bandpass, particularly the spot-only model (see Figure~5 of \cite{rackham19_tlseSunlike}), unlike what we observe. Therefore, the transit light source effect seems unlikely as well. 

A possible systematic explanation for the discrepancy in transit depth between our two HST visits is residual red noise present in the data. As mentioned in Section~\ref{subsec:jointBBfit}, the residuals of the broadband fit of the first visit are consistent with pure white noise behavior, while those of the second visit exhibit excess RMS suggestive of red noise still present in the data. We found the same behavior for each visit from the spectroscopic light curve fits as well, as shown in Figure~\ref{fig:AVplots}. We cannot rule out that this red noise is the cause of the discrepant transit depths, despite the two data sets having been reduced and fit consistently. Fortunately, as shown in Figure~\ref{fig:transpec} and discussed previously, this did not affect the shape or relative amplitudes of the transmission spectra of each visit, and only introduced a constant offset between them. Additional high-precision transit observations of HD~219666~b, particularly with the James Webb Space Telescope, would help shed more light on the origin of this offset. 

\subsection{Disagreement on HD~219666~b's orbital period in the literature} \label{subsec:litcomparison}

There is considerable disagreement between measurements of HD 219666 b's orbital parameters in the literature, most critically its orbital period. In fact, the use of inaccurate orbital parameters led to missing HD 219666 b's transit in most of our HST visits. Our updated constraints of HD~219666~b's orbit are based on the most extensive and high-precision set of data among the other literature works, and are shown to well fit each observation, so we recommend using our measurements when planning future observations to avoid similar issues. 

When planning our HST observations, we used the orbital period of \cite{esposito2019} which was the only available reference at the time, and who only had available to them the 2018 TESS observations. The period we derive in this work is 2.54$\sigma$ different, and is instead extremely consistent ($<$0.11$\sigma$) with that of \cite{hellier2019}, which supplemented the 2018 TESS data with WASP-South observations, as well as that of \cite{kokori2023_exoclock}, who used the 2018 and 2020 TESS data. Our period is also strongly inconsistent ($>$80$\sigma$) with those of \cite{patel2022} and \cite{oddo2023}, who derive shorter and longer periods, respectively. \cite{patel2022} used only the 2018 and 2020 TESS data, while \cite{oddo2023} used both of these as well as new CHEOPS observations. 

The CHEOPS observations that \cite{oddo2023} introduced captured a single transit and did not observe the ingress or egress, so it provides limited novel constraining power to the period. Therefore, the source of these discrepancies seems rooted in the fitting of the 2018 and 2020 TESS visits. Interestingly, using the same data, \cite{kokori2023_exoclock} and \cite{patel2022} derive different periods which are inconsistent at $>$60$\sigma$. This highlights the utility of combining several epochs from several different instruments into a self-consistent fit in order to accurately constrain a planet's orbit and ephemeris. 

\section{Conclusions}

In this work, we presented the near-IR (1.1 - 1.6$\mu$m) transmission spectrum of the hot Neptune HD~219666~b observed from two visits with HST/WFC3 G141. Despite an achromatic transit depth offset between the two visits, we derived morphologically similar transmission spectra from each visit, and performed retrieval analyses on the combined spectrum (Figure~\ref{fig:transpec}). Our fiducial retrieval analysis detected the presence of water in HD~219666~b's atmosphere at 2.9-$\sigma$ significance, in addition to a lack of methane. 

Motivated by recent JWST observations which have detected methane in the atmosphere of several Neptunian-sized planets, despite HST/WFC3 nondetections, we also considered an alternate hypothesis that methane is the dominant contributor to our observed spectrum. We found that a methane-dominated, water-free model can adequately fit our observations (Figure~\ref{fig:loocvresults}), though is physically unlikely and not preferred by Bayesian evidence. We performed a LOO-CV analysis to explore this model preference, finding that the peak of the absorption feature at 1.4~$\mu$m primarily drives our water detection. We do not suggest that HD~219666~b's atmosphere is water-free, but rather highlight the challenge of distinguishing these two molecules using HST/WFC3 G141 data alone in light of recent JWST detections of methane on similar planets. We speculated on what this implies for our population-level understanding of exoplanetary atmospheres, which to date have largely relied on such HST/WFC3 G141 measurements. We found tentative evidence that HST/WFC3 G141-only analyses have systematically underestimated both the water and methane abundances, and therefore the overall metallicities, of Neptunian-sized exoplanets. A more detailed and complete follow-up analysis is necessary to confirm this speculation.

\section*{Acknowledgements}

We thank our reviewer, Dr. Quentin Changeat, for their helpful comments and suggestions that improved our manuscript. We would also like to thank Dr. Everett Schlawin for providing helpful feedback on a draft of our manuscript. M.M.M. acknowledges funding from NASA Goddard Spaceflight Center via NASA contract NAS5-02105. L.W. thanks the Heising-Simons Foundation for their funding through the 51 Pegasi b Postdoctoral Fellowship. L.W. thanks Peter McGill for helpful conversations on LOO-CV. 

This work benefited from the 2024 Exoplanet Summer Program in the Other Worlds Laboratory (OWL) at the University of California, Santa Cruz, a program funded by the Heising-Simons Foundation and NASA. This research made use of the open source Python package exoctk, the Exoplanet Characterization Toolkit \citep{ExoCTK}; the Astrophysics Data System, funded by NASA under Cooperative Agreement 80NSSC21M00561; as well as the NASA Exoplanet Archive, which is operated by the California Institute of Technology, under contract with the National Aeronautics and Space Administration under the Exoplanet Exploration Program.

\vspace{5mm}
\facilities{TESS, HST (WFC3), Exoplanet Archive}
\software{
        astropy \citep{astropy:2013, astropy:2018, astropy:2022}
        batman \citep{kreidberg2015batman},    
        emcee \citep{emcee},
        ExoCTK \citep{ExoCTK},
        matplotlib \citep{matplotlib},
        NumPy \citep{numpy},        
        SciPy \citep{scipy}
          }

\clearpage
\appendix 
\restartappendixnumbering
\section{Additional figures}

\begin{figure*}[h!]
    \centering
    \includegraphics[width=\textwidth]{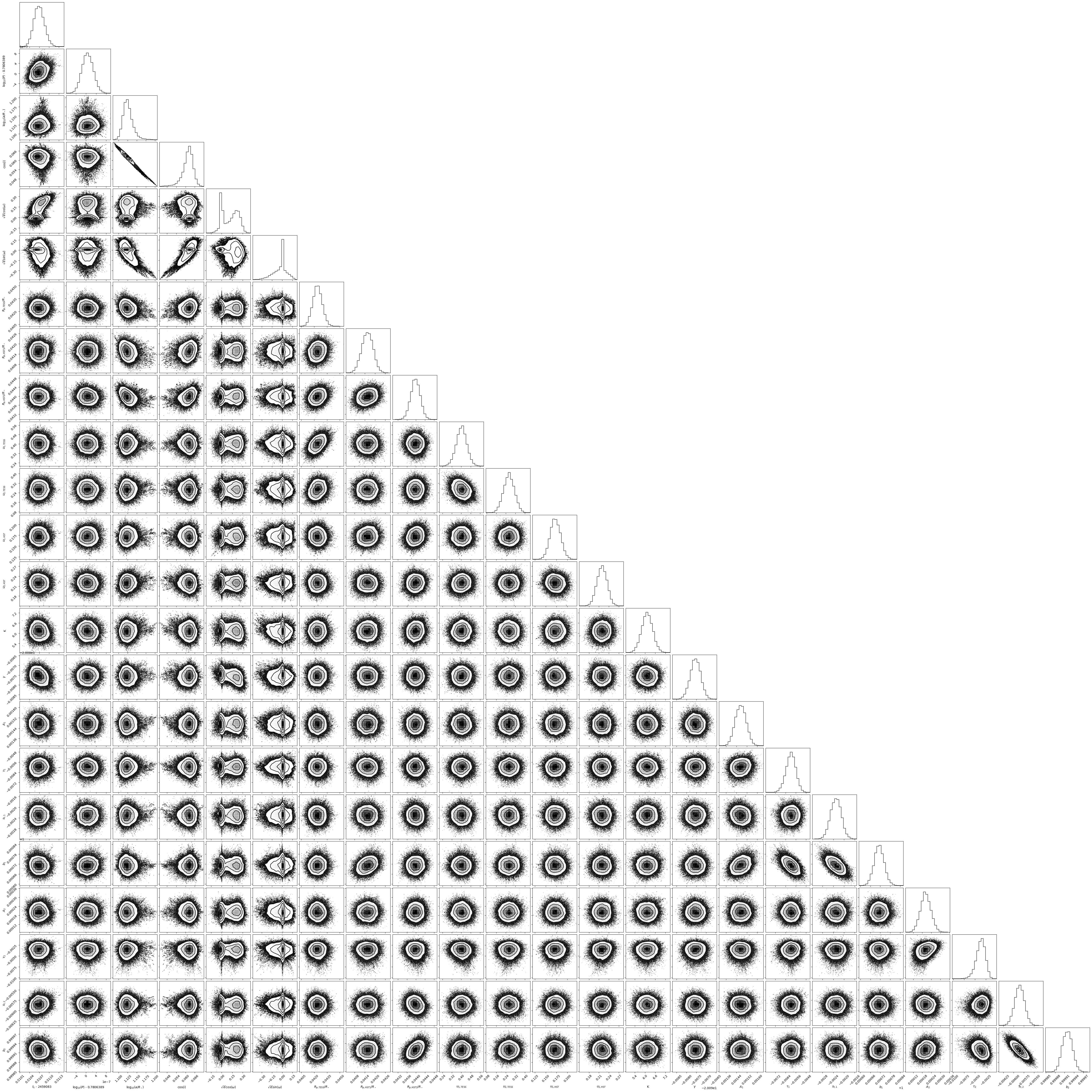}
    \caption{Corner plot from our joint transit and RV fit described in Section~\ref{subsec:methods_LCfitting}.}
    \label{apxfix:jfcornerplot}
\end{figure*}

\begin{figure*}[h!]
    \centering
    \includegraphics[width=\textwidth]{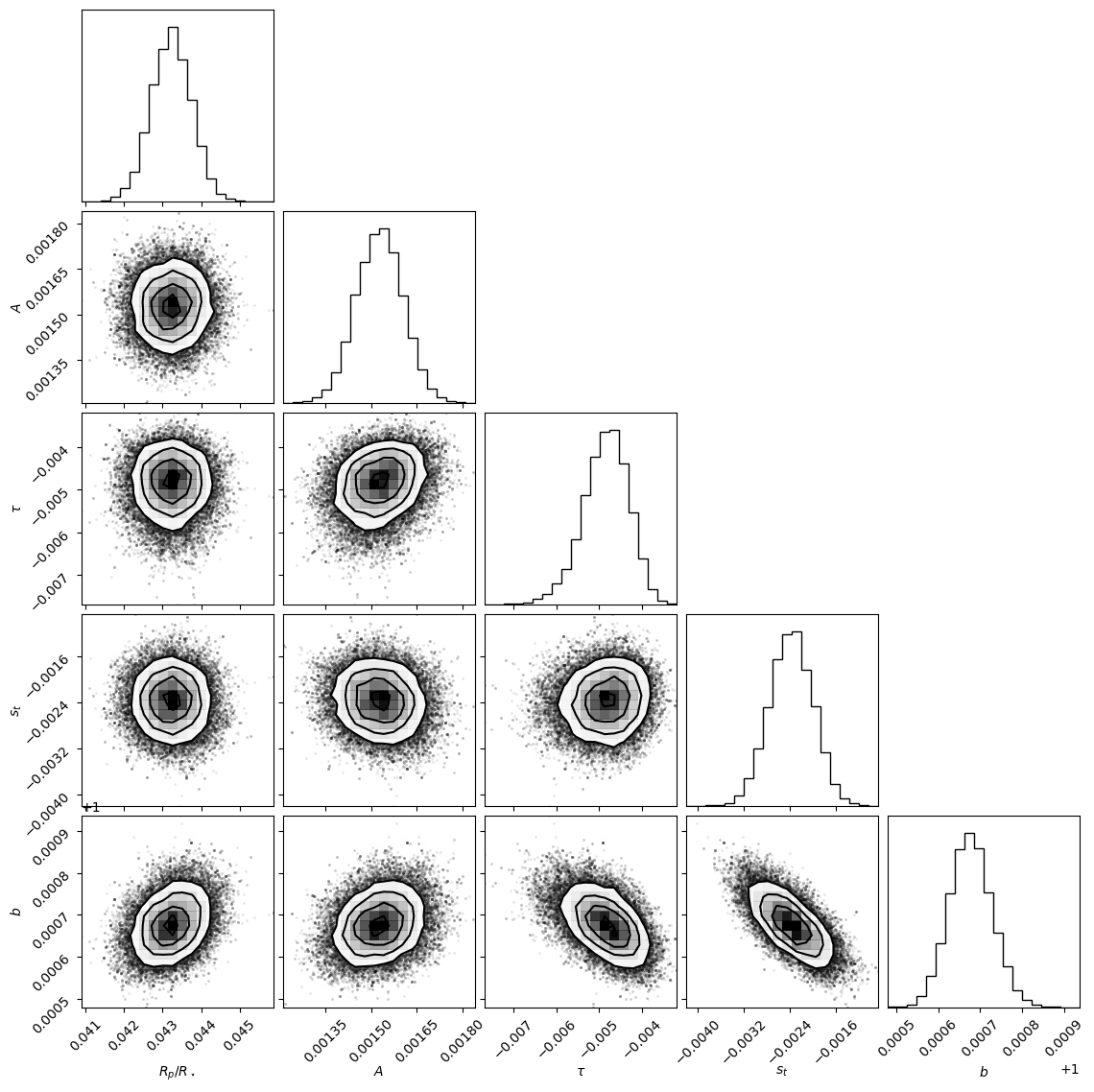}
    \caption{Corner plot from our fit to the spectroscopic HST/WFC3 G141 light curve at $\sim$1.40~$\mu$m from Visit 1.}
    \label{apxfig:speccornerplot1}
\end{figure*}

\begin{figure*}[h!]
    \centering
    \includegraphics[width=\textwidth]{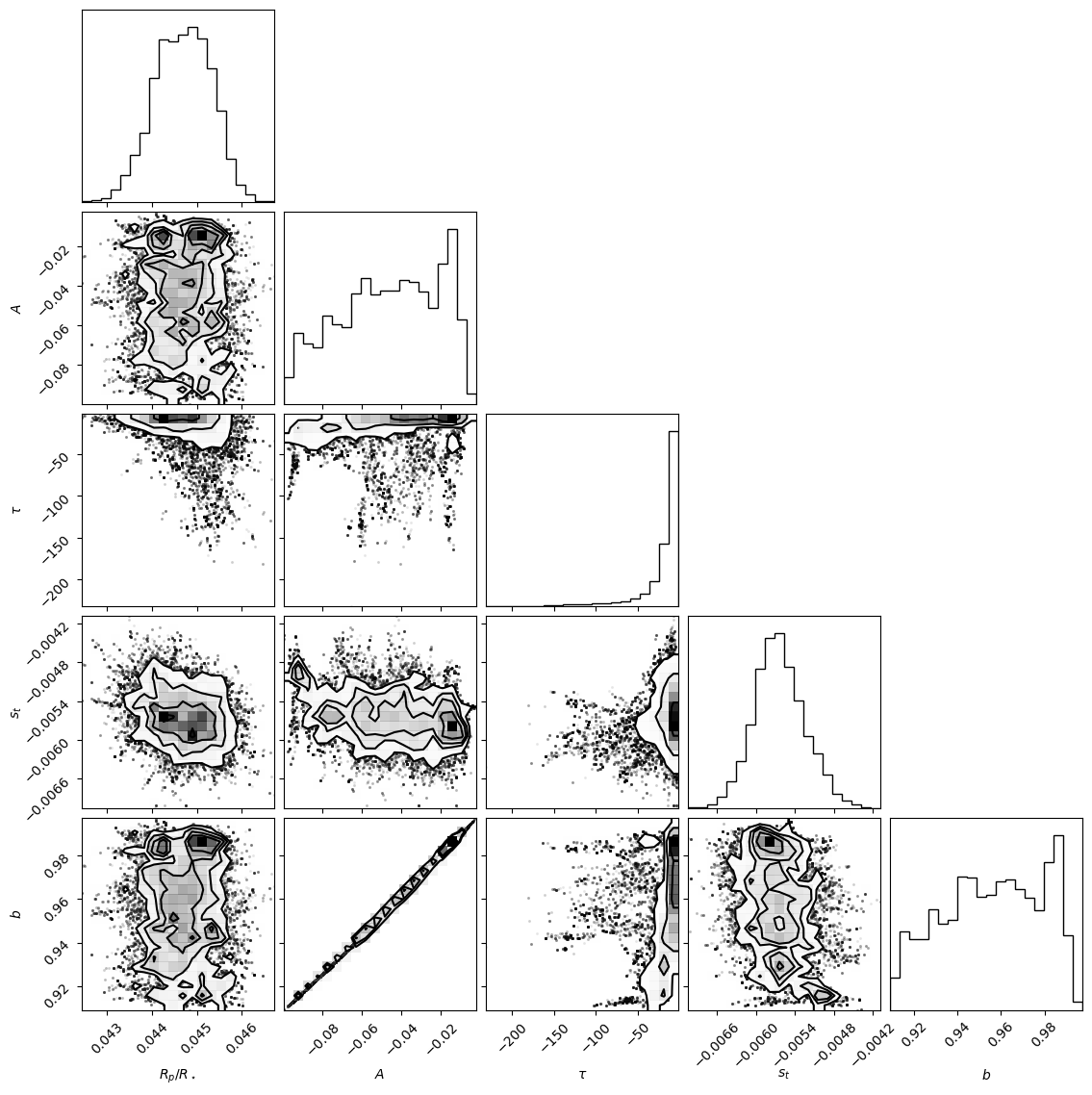}
    \caption{Corner plot from our fit to the spectroscopic HST/WFC3 G141 light curve at $\sim$1.40~$\mu$m from Visit 2.}
    \label{apxfig:speccornerplot2}
\end{figure*}

\begin{figure*}[h!]
    \centering
    \includegraphics[width=\textwidth]{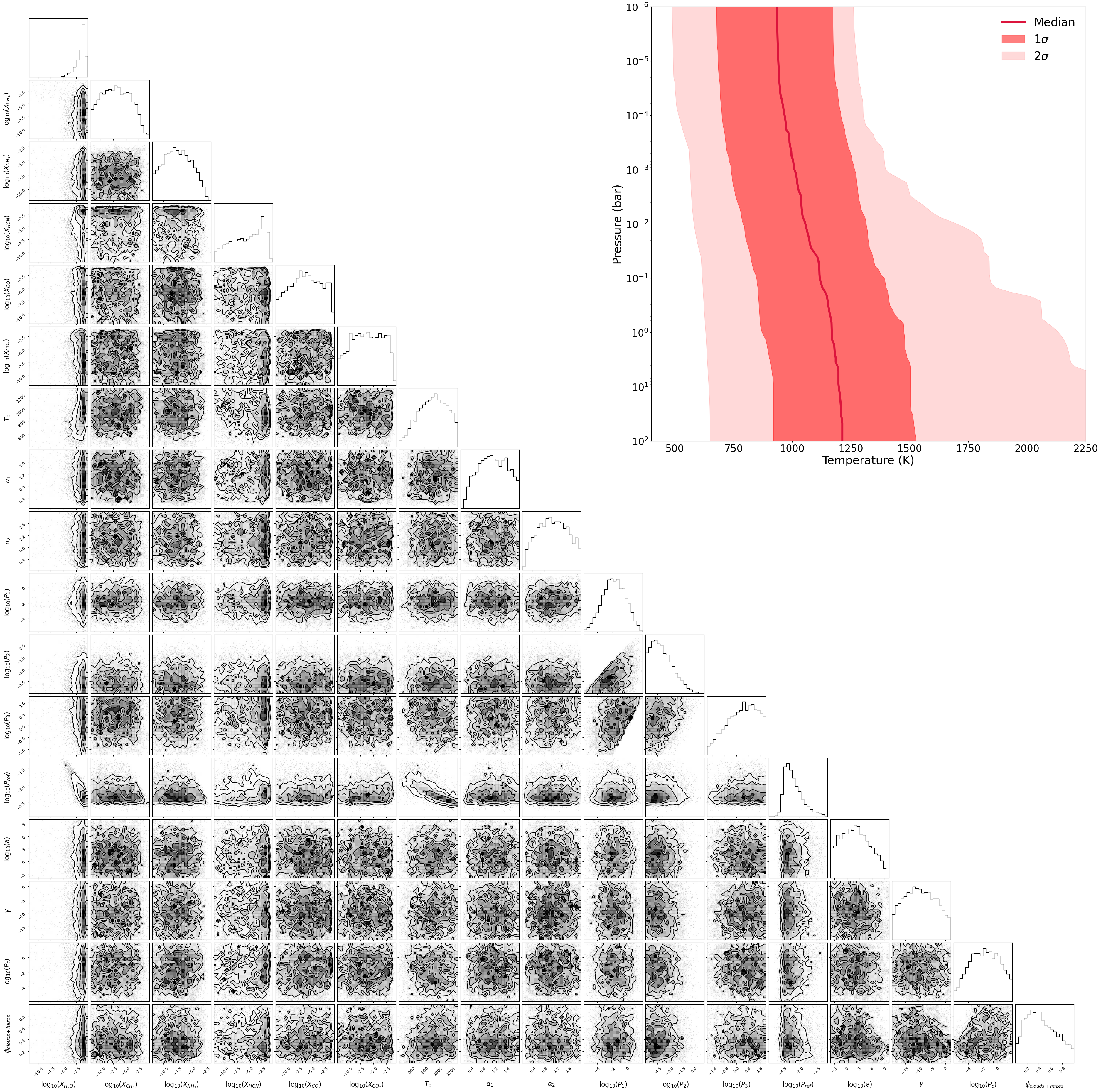}
    \caption{Corner plot from our fiducial atmospheric retrieval on the HST/WFC3 G141 transmission spectrum. Overlaid in the top right of the figure is the retrieved temperature-pressure profile, showing the median profile and 1 and 2$\sigma$ uncertainties.}
    \label{apxfig:fiducialretrievalcornerplot}
\end{figure*}

\begin{figure*}[h!]
    \centering
    \includegraphics[width=\textwidth]{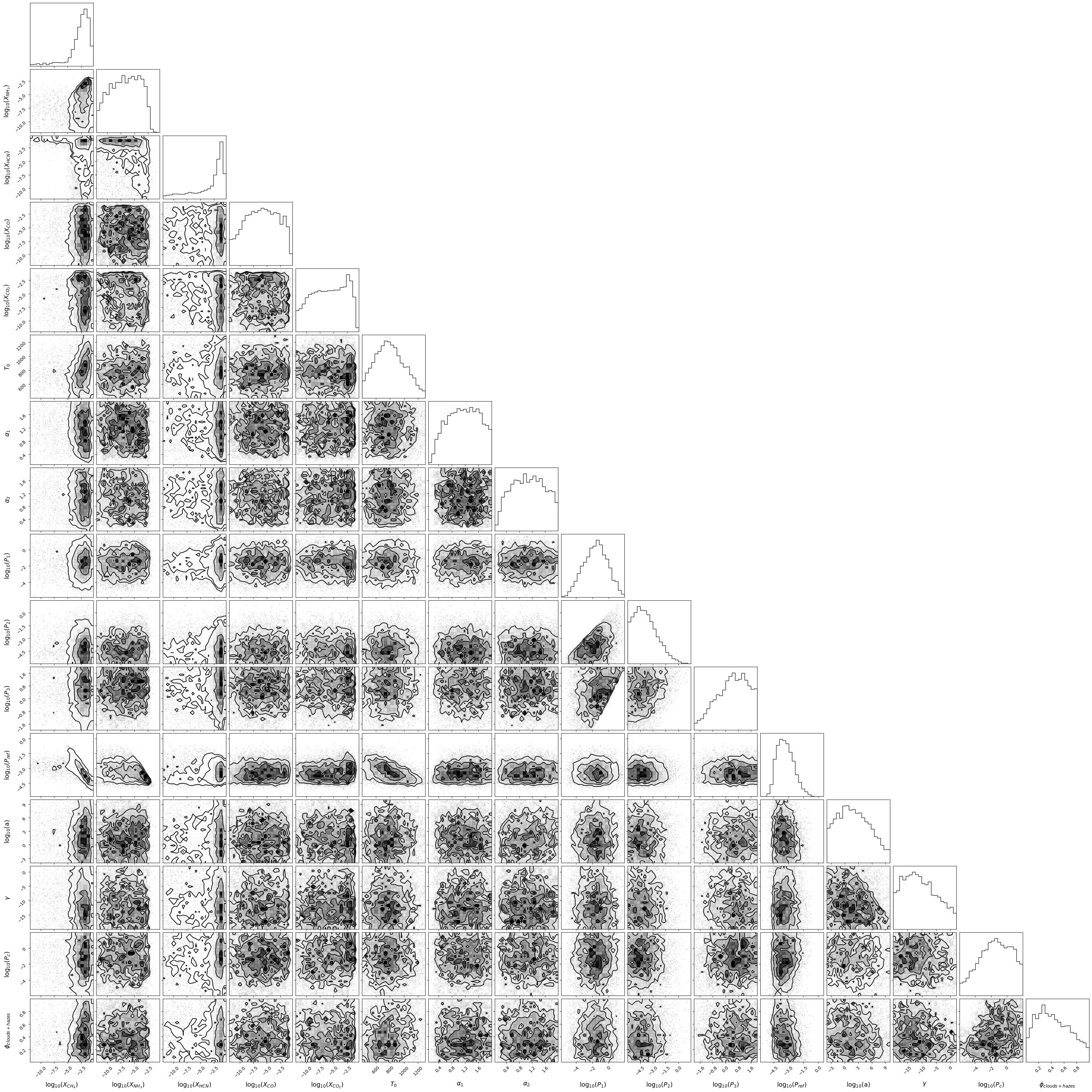}
    \caption{Corner plot from our atmospheric retrieval excluding H$_2$O on the HST/WFC3 G141 transmission spectrum.}
    \label{apxfig:noH2Oretrievalcornerplot}
\end{figure*}

Here we provide additional figures from our analysis. Figures~\ref{apxfix:jfcornerplot}-\ref{apxfig:speccornerplot2} show corner plots from various light curve fits, and Figures \ref{apxfig:fiducialretrievalcornerplot} and \ref{apxfig:noH2Oretrievalcornerplot} show corner plots from various atmospheric retrievals.

\newpage \clearpage

\bibliographystyle{aasjournal}
\bibliography{ref}

\end{document}